\documentclass[usenatbib]{mn2e} 
\usepackage{graphicx,subfigure,times} 

\def\R200{\ensuremath{R_{\mathrm{200}\ }}}

\newcommand{\Zsol}{\ensuremath{\mathrm{Z_{\odot}}}}

\newcommand{\Chandra}{\emph{Chandra}\ }
\newcommand{\Einstein}{\emph{Einstein}\ }
\newcommand{\ROSAT}{\emph{ROSAT}\ }

\newcommand{\MEKAL}{\textsc{MeKaL}\ }

\newcommand{\CIAO}{\emph{CIAO}\ }

\newcommand{\chisq}{\ensuremath{\chi^2}\ }

\newcommand{\nm}{\mbox{\ensuremath{\mathrm{~\nm}\ }}}
\newcommand{\cm}{\mbox{\ensuremath{\mathrm{~cm}\ }}}

\newcommand{\km}{\mbox{\ensuremath{\mathrm{~km}\ }}}

\newcommand{\kpc}{\mbox{\ensuremath{\mathrm{~kpc}}}}
\newcommand{\Mpc}{\mbox{\ensuremath{\mathrm{~Mpc}}}}
\newcommand{\s}{\mbox{\ensuremath{\mathrm{~s}\ }}}

\newcommand{\keV}{\mbox{\ensuremath{\mathrm{~keV}\ }}}
\newcommand{\erg}{\mbox{\ensuremath{\mathrm{~erg}\ }}}

\newcommand{\GHz}{\mbox{\ensuremath{\mathrm{~GHz}\ }}}

\newcommand{\pcc}{\ensuremath{\mathrm{\cm^{-3}\ }}}

\newcommand{\pMpc}{\ensuremath{\mathrm{\Mpc^{-1}\ }}}
\newcommand{\ps}{\ensuremath{\mathrm{\s^{-1}\ }}}

\newcommand{\ergps}{\ensuremath{\mathrm{\erg \ps}\ }}

\newcommand{\kmps}{\ensuremath{\mathrm{\km \ps}\ }}

\newcommand{\To}{--}

\title[X-ray Radio Interactions in Galaxy Clusters]{Interactions of Radio Galaxies and the Intra-Cluster Medium in Abell 160 and Abell 2462}
\author[N. N. Jetha et al.]{Nazirah N. Jetha${^1}$\thanks{Email: nnj@star.sr.bham.ac.uk}, Irini Sakelliou${^1}$, Martin J. Hardcastle${^{2,3}}$, \newauthor Trevor J. Ponman${^1}$ and Ian R. Stevens${^1}$\\
${^1}${School of Physics and Astronomy, University of Birmingham, Edgbaston, Birmingham B15 2TT}\\
${^2}${School of Physics, Astronomy and Mathematics, University of Hertfordshire, College Lane, Hatfield, Hertfordshire AL10 9AB}\\
${^3}${Department of Physics, University of Bristol, Tyndall Avenue, Bristol BS8 1TL}}

\begin{document}

\label{firstpage}

\maketitle

\begin{abstract}
We present \Chandra and VLA observations of two galaxy clusters, Abell~160 and Abell~2462, whose brightest cluster galaxies (BCGs) host wide angle tailed radio galaxies (WATs).
We search for evidence of interactions between the radio emission and
the hot, X-ray emitting gas, and test various jet termination
models.  We find that both clusters have cool BCGs at the cluster centre, and that the
scale of these cores (\(\sim\)30-40\kpc \(\ \)for both sources) is of approximately the same scale as the
length of the radio jets.  For both sources, the jet flaring point is coincident with a steepening in the host cluster's temperature gradient, and similar results are found for 3C465 and Hydra A.  However, none of the published models of WAT formation offer a satisfactory explanation as to why this may be the case.  Therefore it is unclear what causes the sudden transition between the jet and the plume.  Without accurate modelling, we cannot ascertain whether the steepening of the temperature gradient is the main cause of the transition, or merely a tracer of an underlying process. 
\end{abstract}

\begin{keywords}galaxies: active - galaxies: clusters: individual Abell 160 - galaxies: clusters: individual Abell 2462 - X-rays: galaxies: clusters - radio continuum: galaxies  \end{keywords}

\section{Introduction}
\label{introduction}
Wide angle tailed radio galaxies (WATs) are objects located at or near
cluster centres, with long plumes that are often bent, and are generally
hosted by the dominant cluster galaxy.  However, not all radio galaxies with
bent plumes meet our definition of a WAT.  We follow the definition of
\citet{Leahy93}, in which WATs initially have well collimated,\kpc-scale jets which suddenly flare into diffuse plumes, which may be significantly bent.

WATs have traditionally been classified as a subset of Fanaroff and
Riley type I galaxies \citep{fr1974} because of their large-scale
structure.  Unlike classical FRIs, which have jets with very
wide opening angles, WATs have narrow collimated jets on small
scales (tens of \kpc) that resemble those of FRII
sources.  The disruption of WAT jets occurs at the base of the plumes, rather than at the end of the sources, as is the case for FRII sources \citep{HS2003}.

The large scale plumes in WATs can be shaped by the interactions with
their environments. For example, plumes can be bent backwards due to the
relative motion of the host galaxy through the intra-cluster medium (ICM) \citep{BRB79}.  The galaxy is not at rest with respect to the ICM, as one would expect of the brightest cluster galaxy (BCG) in
a virialized cluster, but rather has some (small) velocity.
\citet{Gomez97} interpret this as a sign of some cluster-cluster
interaction or merger, which disturbs the cluster potential.  However, many  WAT sources are
not significantly bent, so that it has been argued that the plume bending is not related to the characteristic jet flaring.  For instance, 0110+152 exhibits sudden jet flaring, but has relatively straight plumes; whereas other sources,
such as 3C75 \citep[see, for example][]{HS2003}, exhibit large scale bending, but not the sudden jet flaring.  Attempts to explain the large scale bending have been explored in detail elsewhere (eg \citealt{1994AJ....108.2031P}, and \citealt{1993ApJ...408..428O}).  In this paper, we address the possible explanations for the sudden jet flaring.

\begin{table*}
\caption{Cluster and WAT characteristics}
\begin{tabular}{cccccccccc}
\hline& \multicolumn{2}{c}{Cluster centre coordinates}& & && && \multicolumn{2}{c}{Radio core coordinates}\\ 
Abell Cluster & \(\alpha_{2000}\)& \(\delta_{2000}\) &\(z\) & scale & \(\mathrm{D_L}\) & R\(_{500}\)&Radio source name &\(\alpha_{2000}\)&\(\delta_{2000}\)\\
              &                  &                   &  & arcsec/\kpc&(\Mpc)&(\kpc)\\ \hline 

Abell~160 &  01 13 03.9\(^a\) & +15 29 43.9 & 0.0447& 0.88 & 180& 870&0110+152 & 01 12 59.6 & +15 29 28.7\\ 
Abell~2462 & 22 39 05.2\(^b\) & --17 19 53.6& 0.0737& 1.57 & 320& 930&2236--176& 22 39 11.4 & --17 20 27.3 \\\hline
\end{tabular}
\\
\vspace{0.2cm}
\begin{minipage}{16cm}
\small NOTES:\\ 
\(a\): coordinates taken from \citet{Acreman}.\\
\(b\): coordinates taken from \citet{Gomez97}.\\
\(D_L\) is the luminosity distance to each cluster.
\(R_{500}\) is the radius within which the mean density of the cluster is equal to 500 times the critical density of the universe.
\end{minipage} 
\label{clusters}
\end{table*}

It is thought that the denser cluster environment is responsible for the characteristic jet-plume transition of WATs, with the cluster properties determining the location of the base of the plume; WATs are found exclusively in cluster environments, whilst FRIIs of similar radio powers avoid such environments.  X-rays are the best source of information regarding the state of the hot ICM, so using long exposure, high resolution observations with \Chandra\(\!\), we can probe the state of the ICM, and determine which if any of the models proposed to explain the sudden jet flaring fit the observed data.

In this paper, we present \Chandra observations of the two nearby
galaxy clusters Abell~160, and Abell~2462, (for details of cluster properties see Table~\ref{clusters}).  The central cluster galaxies of both these clusters are WAT hosts.  Abell~160 is a poor cluster (richness class 0), and was first observed with the {\sc{ipc}} instrument on board the \Einstein satellite.  This showed emission from the cD galaxy, as well as some evidence of extended emission from the cluster \citep{1999ApJ...511...65J}.  X-ray observations with \ROSAT showed extended
cluster emission that was fairly uniform in structure \citep{Drake}.  However, the temperature of the cluster has not been previously measured by either \Einstein or \ROSAT\@.  Abell~160 hosts the WAT 0110+152.  Abell~2462 is another poor Abell
cluster.  \citet{Gomez97} working with \ROSAT data suggested that the X-ray structure was somewhat irregular, and that some elongation existed from the centre of the cluster towards the east.  They found a temperature for Abell~2462 of 1.5\(^{+0.8}_{-0.2}\)\keV\(\!\).  Abell~2462 hosts the WAT 2236-176.

In Section~\ref{xrayobs}, we present the new \Chandra observations, and in
Section~\ref{radioobs}, the radio observations.  We describe the X-ray spatial and
spectral analysis in Section~\ref{xrayanal}.  In Section~\ref{discussion}, we discuss our results
and the possible explanations in the light of current models (discussed
above) for jet disruption and termination.  

We assume \(H_0\)=70\kmps\( \! \!\)\pMpc\(\!\) and use J2000 co-ordinates throughout.

\section{X-Ray Observations}
\label{xrayobs}

The clusters Abell~160 and Abell~2462 were observed with \Chandra on
2002 October 18 and 2002 November 19, for $\sim$60 and $\sim$40~ksec
respectively.  During both observations, the detectors were operating
in the {\sc{vfaint}} mode.  The prime instrument for the Abell~160
observation was {\sc{acis-i}}, whilst the pointing for Abell~2462 was
centred on the {\sc{s3}} chip of {\sc{acis-s}}.  Information on the \Chandra observations is given in Table \ref{pointings}.

The data were reprocessed using {\sc{ciao}} version 2.3 and
{\sc{caldb}} 2.23, to use the {\sc{vfaint}} mode cleaning techniques.
Additionally, the data were corrected for the {\sc{acis}}
charge transfer inefficiency (CTI) using the
{\sc{acis\_process\_events}} script.

Light curves in the energy range 0.3\To10.0\keV were created for each
data set, in regions outside the cores of both clusters, showing that neither
of the observations were severely contaminated by background flares.
Subsequently, they were subjected to a 3\(\sigma\) clipping, using the
usual {\sc{ciao}} script, {\sc{analyze\_lt.sl}}.  The unfiltered and
corrected (for periods of high background contamination) exposure
times are shown in Table~\ref{pointings}.

We used the {\sc ciao} script {\sc{merge\_all}} to create exposure maps, at
an energy of 1.5\keV, for both clusters.  For any subsequent
spatial analysis (e.g. Section~4.1), we divided the cluster images by
the exposure maps to correct for exposure effects.

A blank-sky background file was also created for Abell~160, by
following the relevant {\sc{ciao}} thread.  This was used to obtain the radial background profiles in the subsequent analysis of Abell~160 (Section~\ref{deprojanalys}), as the cluster fills the {\sc{acis-i}} field of
view, and so finding areas devoid of cluster emission was difficult.
The blank sky files were subjected to the same cleaning and
reprocessing methods as the data events file.

\begin{figure*}
\subfigure{\scalebox{.4}{\label{images:a}\includegraphics{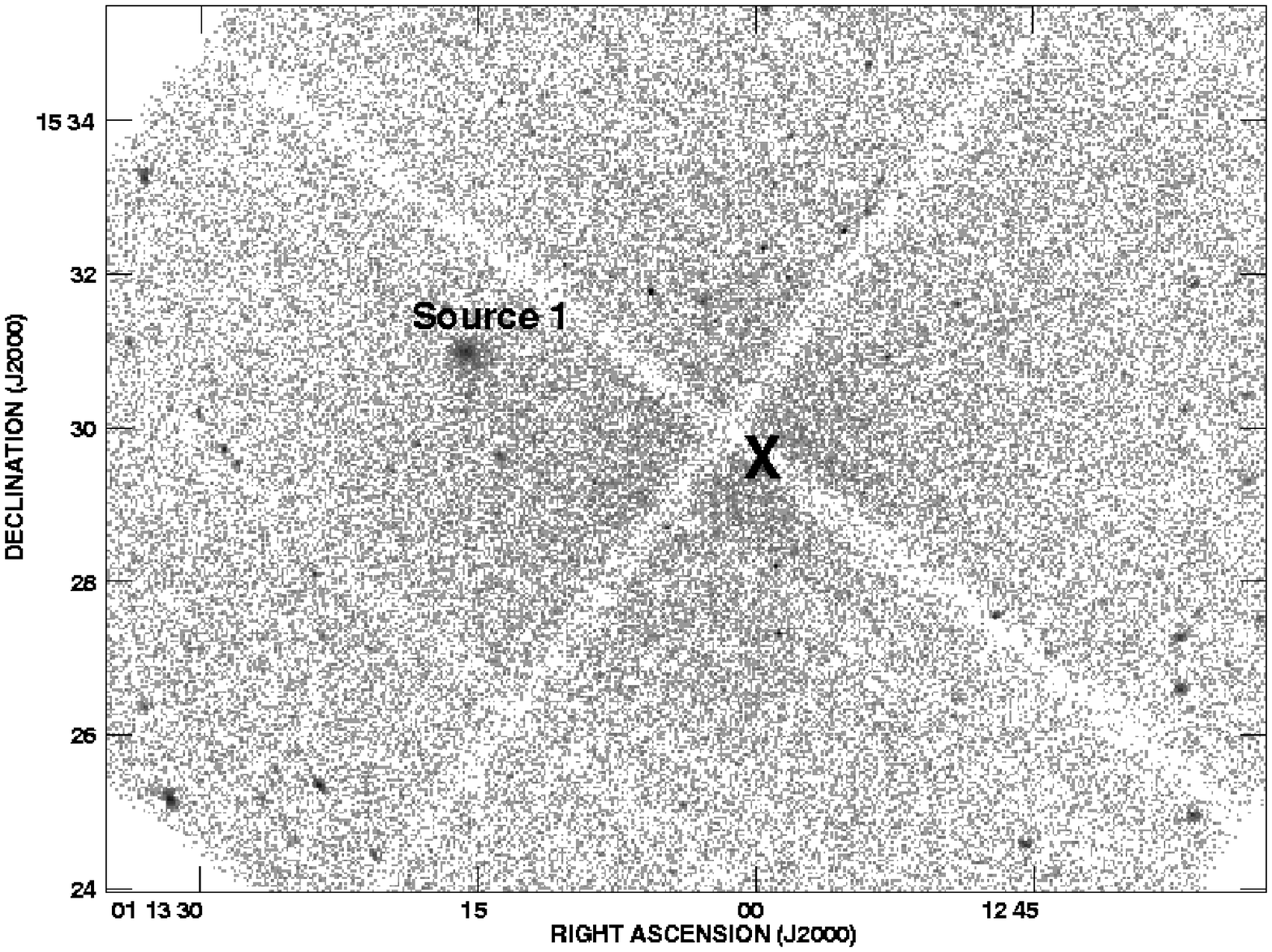}}}
\subfigure{\scalebox{.42}{\label{images:b}\includegraphics{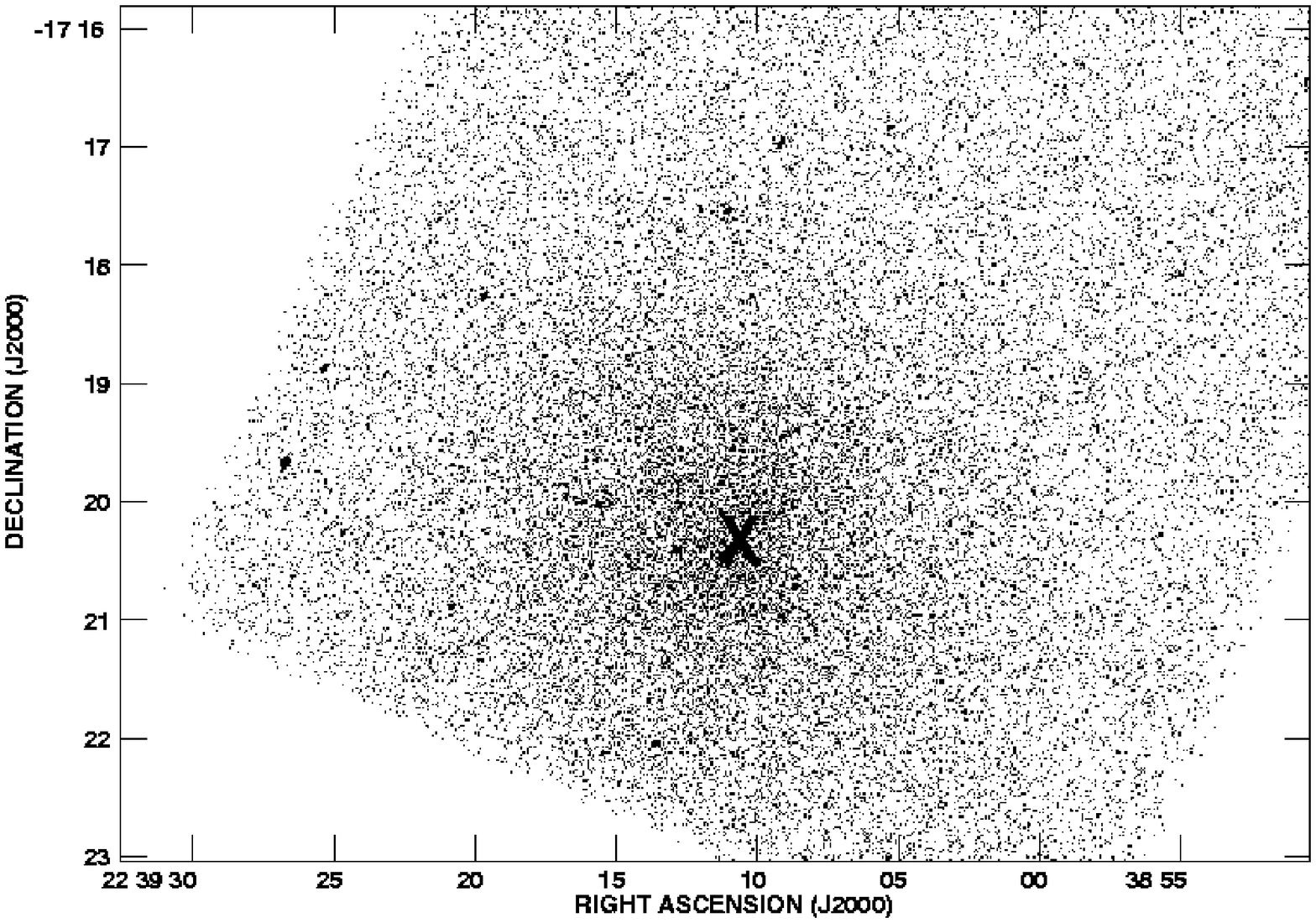}}}
\caption{\Chandra images in the energy range 0.3-5.0\keV\(\!\) of Abell~160 (left panel) and Abell~2462 (right panel).  The position of the BCG in each cluster is marked with a cross.}
\label{images}
\end{figure*}

Raw count images of both clusters in the 0.3--5.0\keV
energy range are shown in Fig.~\ref{images}.  The pixel size is
1.96 by 1.96~arcsec.  We note that on large scales, the emission appears to be fairly symmetric and uniform.  The bright, extended source in Abell~160 at \( \alpha_{2000}\)= 1\(^{h}\)13\(^{m}\)15\(^{s}\)
\(\delta_{2000}\)=+15\(^{o}\)30\(^{\prime}\)57\(^{\prime\prime}\) is labelled 'Source~1' in Fig.~\ref{images}, and we investigate its properties in Section~\ref{source1}.  In Fig.~\ref{radioimages}, we present a smoothed X-ray image of the cluster core, in the same energy range as the raw counts image.  The X-ray images have been smoothed with a Gaussian beam of width 3~arcsec and are overlaid with the radio contours described in Section~\ref{radioobs}.

\begin{table}
\label{pointings}
\caption{Chandra observations} 
\begin{tabular}{cccc}
\hline
Source & Detector\(^a\) & Unfiltered & Corrected\\
       &          & Exposure & Exposure \\
       &          & (ks)     & (ks)\\ \hline 
Abell~160 &{\sc{acis-i}} & 58.50 & 57.49\\ Abell~2462
&{\sc{acis-s}}(S3) & 39.24 & 38.24\\\hline
\end{tabular}
\vspace{0.2cm}
\begin{minipage}{6.5cm}
\small NOTES:\\
\(a\): `Detector' refers to the chip and/or chip array that was used as the prime instrument for each observation.\\
\end{minipage} 
\end{table}

\section{Radio Observations}
\label{radioobs}

VLA data at 1.4\GHz were obtained from the VLA archive for Abell~160.
The data were taken in the A and C array configurations and were
originally presented in \citet{ODonoghue}, together with relevant
polarization maps.  The archival data were calibrated using 3C48 and 0019-000 before {\sc clean} maps were made using the {\sc aips} task {\sc imagr}.  The Abell~2462 radio data were originally presented in \citet{HS2003},
together with polarization maps. The radio source was observed with
the VLA at 8.4\GHz in all four array configurations, and the data were processed as described in \citet{HS2003}.  

\begin{figure*}
\subfigure{\scalebox{.42}{\label{radioimages:a}\includegraphics{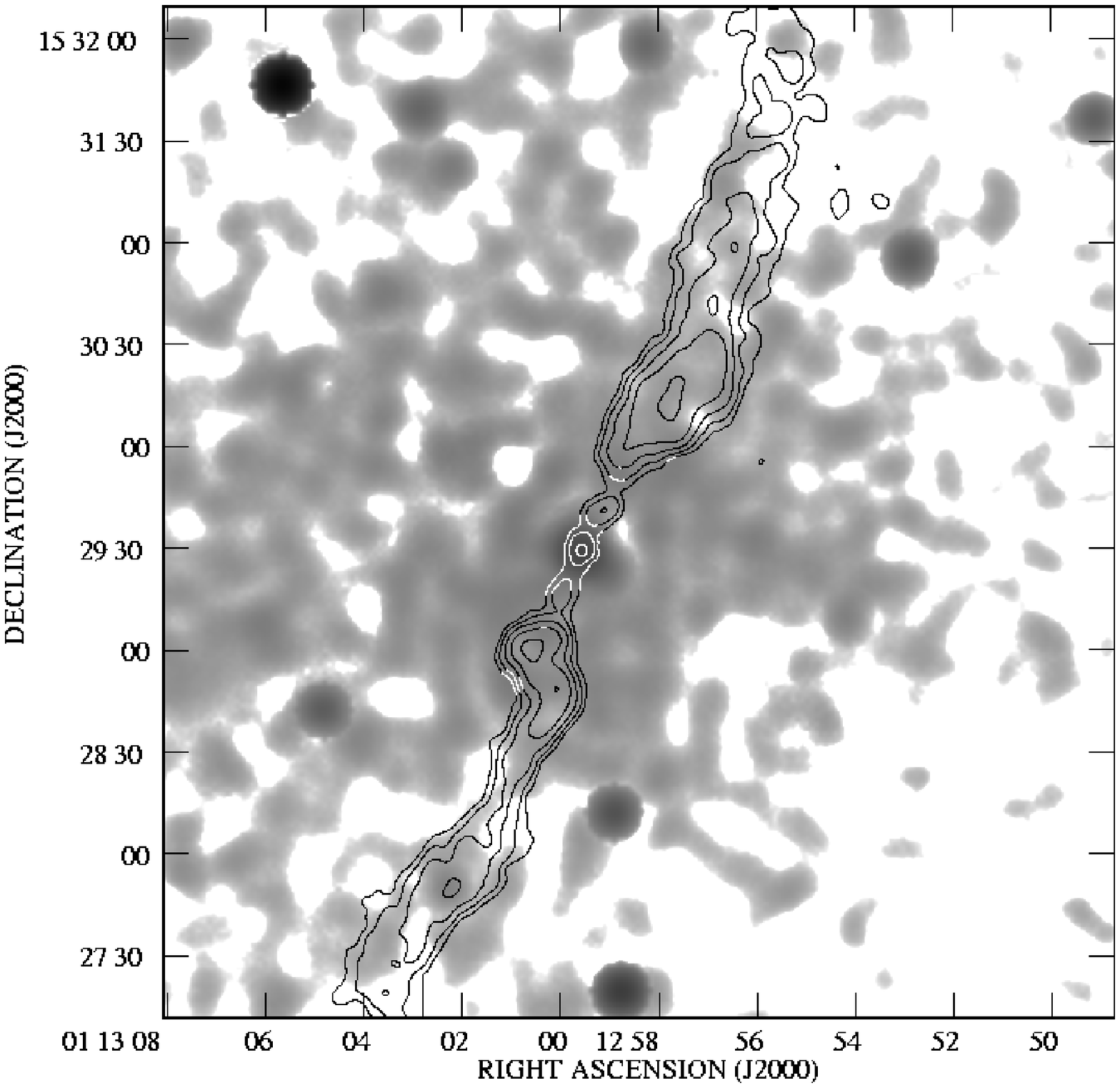}}}
\subfigure{\scalebox{.45}{\label{radioimages:b}\includegraphics{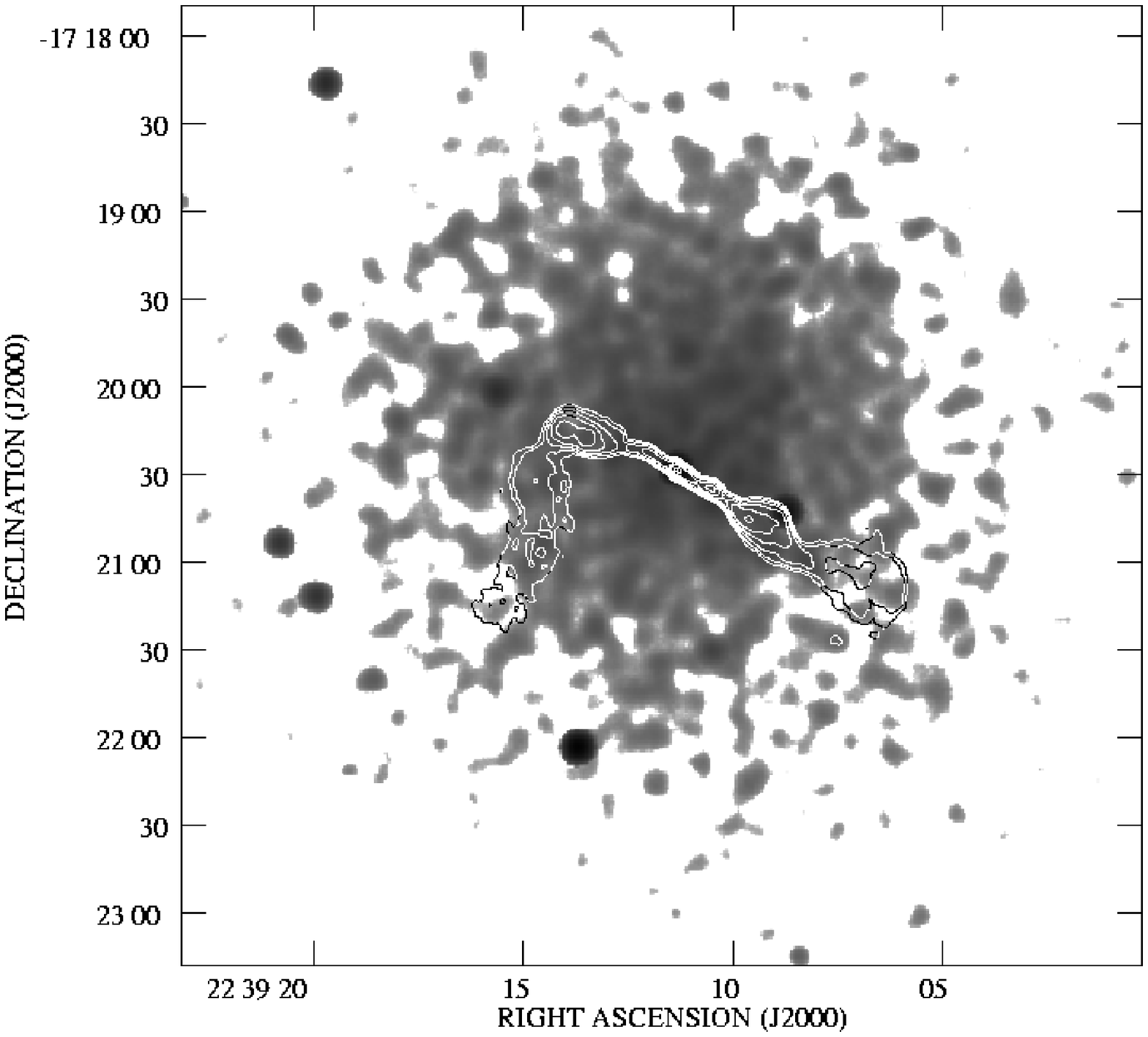}}}
\caption{X-ray images of the central regions of the clusters, with the radio contours overlaid.  Abell~160 is on the left and Abell~2462 on the right.  The resolution of the radio images is  6.4 by 6.4~arcsec for 0110+152 and 0.70 by 0.52~arcsec for 2236-176.  The lowest contours are at 0.94 and 0.18~mJybeam\(^{-1}\) for 0110+152 and 2236-176 respectively, and contours are spaced at 1,2,4,8,16,32 and 64 times the lowest contour values.}
\label{radioimages}
\end{figure*}

We present the radio data in Fig.~\ref{radioimages} as contours overlaid on X-ray images of the central region of each cluster. 

Both sets of data show narrow, well--collimated jets that flare into diffuse straight plumes.  Additionally, the eastern lobe in 2236-176 bends suddenly through 90~degrees, 25~arcsec from the jet flaring point.  Details of the radio source dimensions are given in Table~\ref{radiodimensions}.

\begin{table}
\label{radiodimensions}
\caption{Radio jet properties}
\center  
\begin{tabular}{ccc}
\hline Source & Jet Direction & Jet Length \\
              &               &(arcsec)    \\\hline
0110+152      & NW            & 40\\ 
              & SE            & 40 \\ 
2236-176      & NE            & 28 \\ 
              & SW            & 28  \\\hline
\end{tabular}
\vspace{0.2cm}
\begin{minipage}{5cm}
\small NOTES:\\
The jet lengths are taken from \citet{HS2003}\\
\end{minipage}
\end{table}

\section{X-Ray Data Analysis}
\label{xrayanal}
\subsection{Spatial analysis}
\label{sbanal}

To make an initial estimate of the size of the extended galactic emission, and to map the transition from the galactic to the cluster atmosphere, binned, exposure corrected images (as described in Section \ref{xrayobs}) were
used to extract the 1-D surface brightness profiles for both data
sets.  Point sources were excluded, and for Abell~160, the group
(see Section \ref{xrayobs}) was also excluded.  Both profiles were centred
on the radio core, and adaptively binned so that each bin had a minimum signal to noise ratio of 3.

Profiles were extracted out to 5 (\(\sim\)260\kpc) and 3~arcmin
(\(\sim\)280\kpc) (\(\sim\) 0.3 \(\mathrm{R_{500}}\), see Table~\ref{clusters}) for Abell~160 and Abell~2462 respectively.
Backgrounds were annuli centred on the radio cores, from
5 to 7.7~arcmin for Abell~160 and from 3 to 4~arcmin for Abell~2462.
The profiles were fitted with a composite model comprising of three components -- two \(\beta\) models for the extended galactic and cluster emission, and a point spread function (PSF) to model the emission from the active core.  The \(\beta\) models are standard \(\beta\) models \citep{KFF76}, defined by: 
\begin{equation}
\Sigma\left(r\right)=\Sigma_0\left[1+\left(\frac{r}{r_0}\right)^2\right]^{-3\beta+0.5}
\end{equation} where \(\Sigma\left(r\right)\) is the surface brightness at a distance \(r\) from the centre of the distribution, \(\Sigma_0\) is the central surface brightness, and \(r_0\) is the core radius of the distribution.  The PSF parametrization of 0331+39 from \citet{WBH2001} was used for both sources, since the spectrum of this source was similar to those of both Abell~160 and 2462 (see Section~\ref{specan} for details of how we extracted and fitted the spectra).  

This composite model was fitted to the surface brightness data, leaving the normalizations, core radii, and \(\beta\) values of the two \(\beta\) components, and the normalization of the PSF component free to vary.

\begin{figure*}
\center \subfigure{\scalebox{.4}{\includegraphics{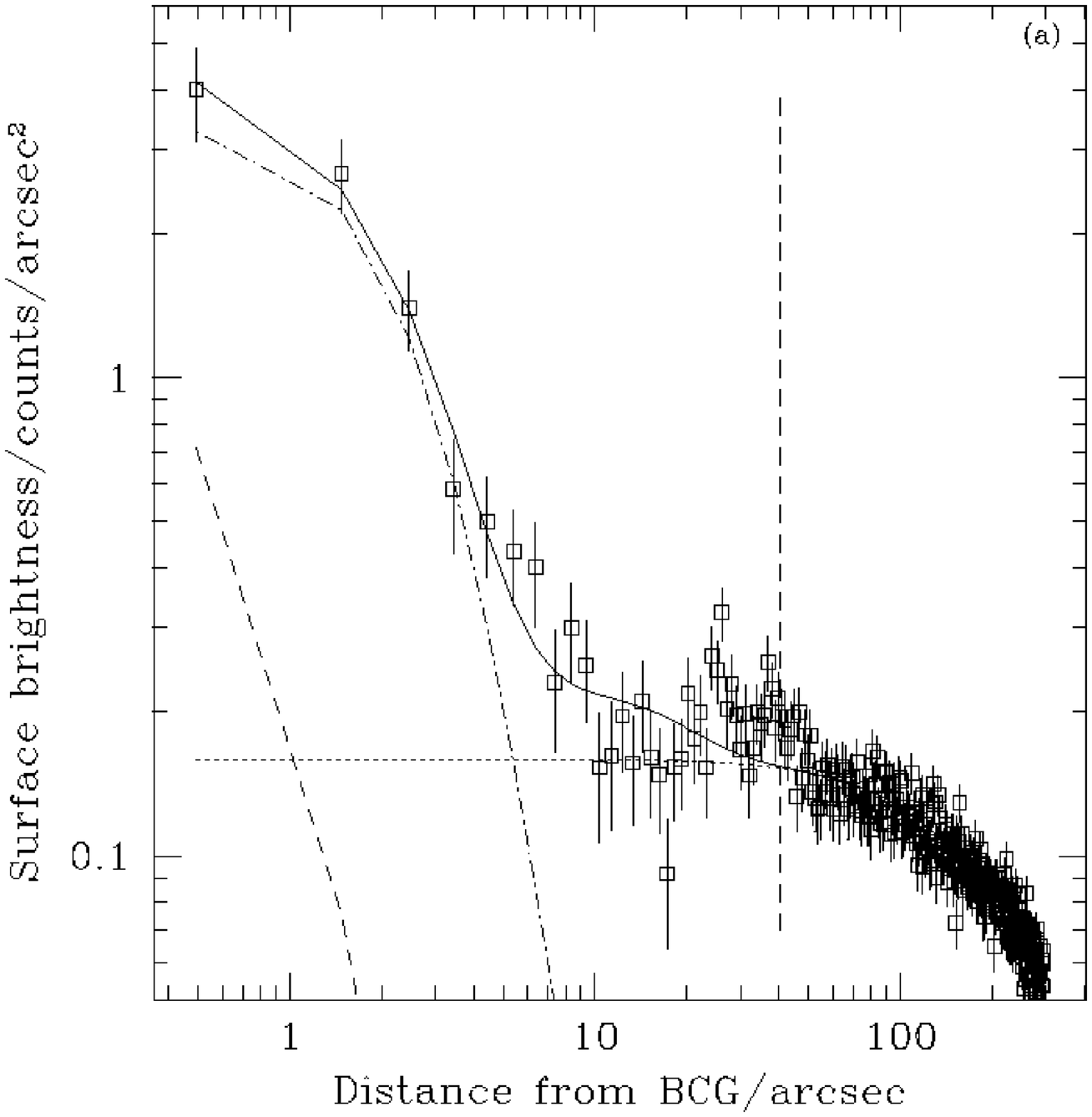}}} 
\subfigure{\scalebox{.4}{\includegraphics{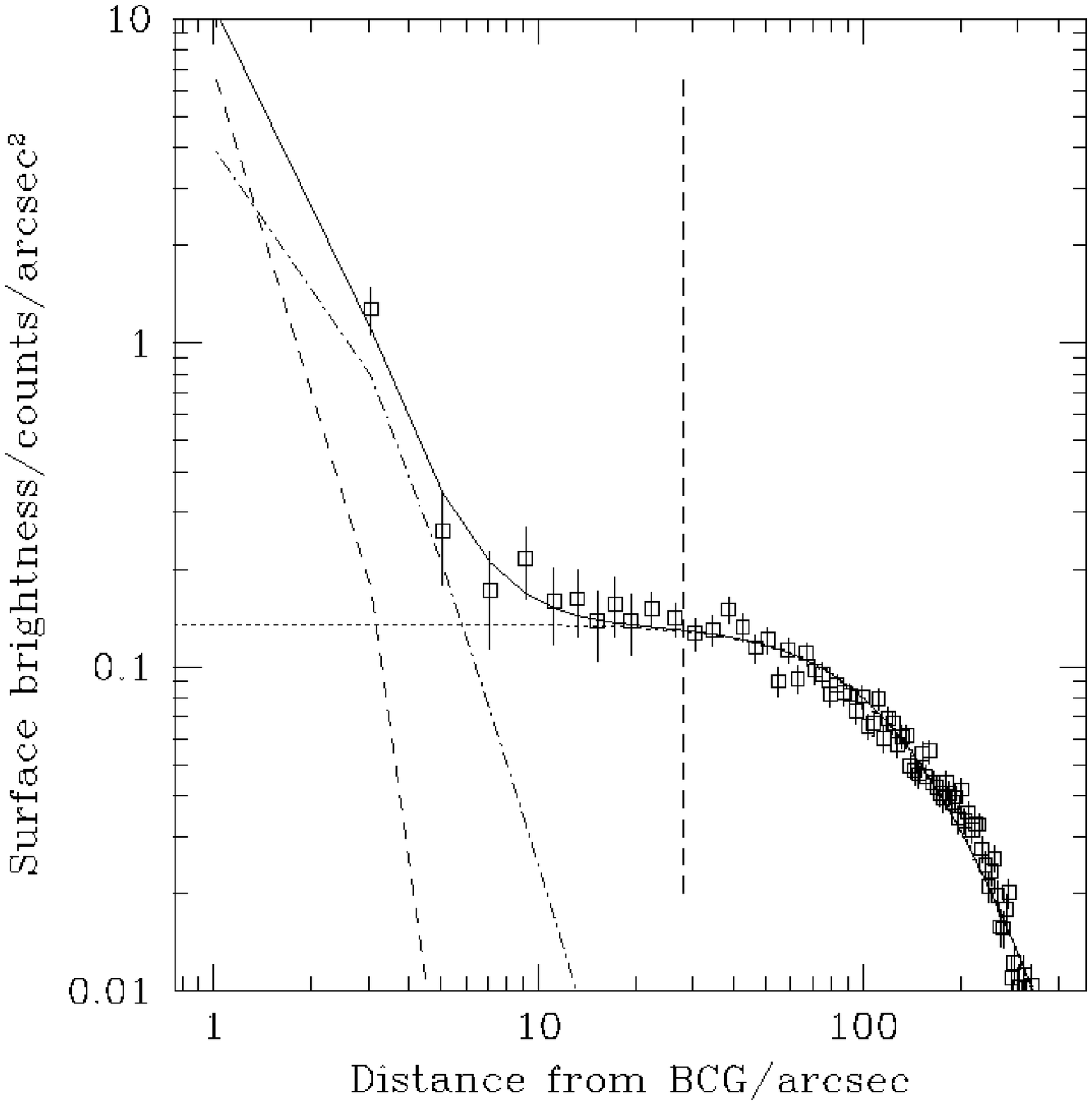}}}
\caption{The 1-D radial surface brightness profiles for both clusters -- figure (a) is the profile for Abell~160, and figure b for Abell~2462. 
The solid line is the composite model, the
dashed line indicates the contribution from the PSF (modelled as
described in the text), the dot-dash line is the contribution from the
steeper \(\beta\) model for the galaxy, and the dotted line is the
contribution modelled by the second \(\beta\) model.  The dashed vertical line shows where the radio jets flare.}
\label{sbprofiles}
\end{figure*}

The profiles and best-fitting models are presented in
Fig.~\ref{sbprofiles}, and the best-fitting parameters are listed in
Table~\ref{sbvalues}.  We find that for both sources the composite model fits the data well.  In both cases, one component is a steep \(\beta\)-model fitting the galactic emission on scales of 10--20~arcsec, while the second flatter component fits the extended cluster-scale emission.  For clusters, \(\beta\) values of \(\sim\)0.6 \citep{Sarazin} are expected, and our data agree well with this.

\begin{table}
\caption{Best-fitting parameter values for the surface brightness fits for
Abell~160 and Abell~2462.  Errors given are 1\(\sigma\) for two interesting
parameters.} 
\begin{tabular}{llll}
\hline\\ Source & Component & Parameter & \\
\hline
A~160        &PSF                & Norm.                   & \(4.0\pm0.01\)      \\
             &                   & (counts/arcsec\(^2\))   &                     \\
             &Galaxy             & \(r_{0_{gal}}\)         &\(3.3^{+1.5}_{-0.9}\)\\
	     &\(\beta\) Model    &(arcsec)	           &                      \\ 
             &                   & \(\beta_{gal}\)         &\(0.96^{+0.5}_{-0.2}\) \\ 
             &                   & Norm.                   & \(3.45^{+0.7}_{-0.8}\) \\ 
             &                   & (counts/arcsec\(^2\))   &                         \\
             &Cluster            & \(r_{0_{clus}}\)        & \(191^{+26}_{-21}\)      \\
             & \(\beta\) Model   & (arcsec)                &                           \\ 
             &                   & \(\beta_{clus}\)        & \(0.53^{+0.06}_{-0.05}\)  \\ 
             &                   & Norm.                   & \(0.16^{+0.004}_{-0.004}\) \\ 
             &                   &(counts/arcsec\(^2\))    &                            \\ 
             &\chisq (d.o.f)    &                         & 391(313) \\\hline

A~2462       &PSF                & Norm.                   & \(20.0\pm0.01\)             \\
             &                   &(counts/arcsec\(^2\))    &                              \\
             &Galaxy             & \(r_{0_{gal}}\)         &\(2.3\pm0.3\)                  \\ 
             & \(\beta\) Model   &(arcsec)                 &                                \\
             &                   & \(\beta_{gal}\)         &\(0.74\pm0.1\)                   \\ 
             &                   & Norm.                   & \(5.6^{+2.0}_{-1.5}\)           \\
             &                   & (counts/arcsec\(^2\))   &                                  \\
             &Cluster            & \(r_{0_{clus}}\)        &\(155^{+1}_{-6}\)            \\ 
             & \(\beta\) Model   &(arcsec)                 &    \\
             &                   & \(\beta_{clus}\)        & \(0.67^{+0.01}_{-0.01}\)  \\ 
             &                   & Norm.                   &\(0.13^{+0.001}_{-0.001}\)  \\
             &                   &(counts/arcsec\(^2\))    &                            \\ 
             &\chisq (d.o.f)    &                         &90 (82) \\\hline

\end{tabular}
\label{sbvalues}
\end{table}

\subsection{Spectral analysis}
\label{specan}

To further investigate the environment in which the jets propagate, and how the environment may relate to the causes of the jet flaring, we investigated the spectral properties of the WAT host clusters (Section~\ref{speccluster}) and the WAT host galaxies (Section~\ref{specgals}).

\subsubsection{The WAT host clusters}
\label{speccluster}

We accumulated counts in circular regions centred on the BCGs out to 5 and 3~arcmin for Abell~160 and Abell~2462 respectively.  The background regions were concentric circular annuli extending from 5--7~arcmin for Abell~160 and 3--4~arcmin for Abell~2462.  From Section~\ref{sbanal}, it was apparent that emission from the WAT host galaxy is dominant out to 10 and 5~arcsec for Abell~160 and Abell~2462 respectively.  The emission from the BCGs was masked out by excluding circular regions, centred on the BGCs, of radius 20 and 10~arcsec (Abell~160 and Abell~2462 respectively).

Spectra were extracted using the {\sc{acisspec}} script, and the
weighted response files were corrected for the degradation in the
quantum efficiency of the {\sc{acis}} chips, using standard \CIAO
tools. We binned the spectra such that each energy bin contained a minimum
of 40 counts after background subtraction.

\begin{figure*}
\subfigure[Abell~160]{\scalebox{.3}{\includegraphics[angle=270]{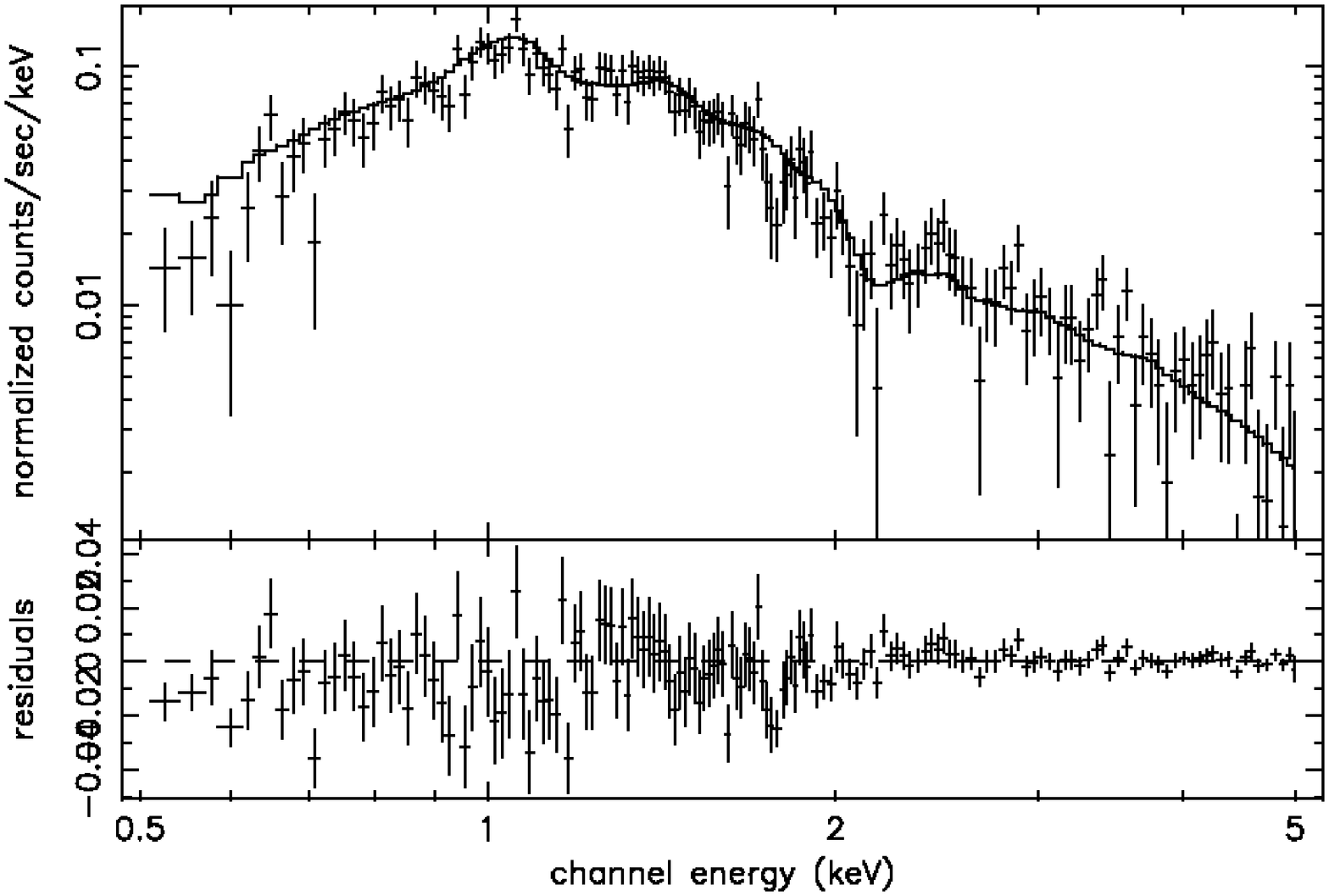}}}
\subfigure[Abell~2462]{\scalebox{.3}{\includegraphics[angle=270]{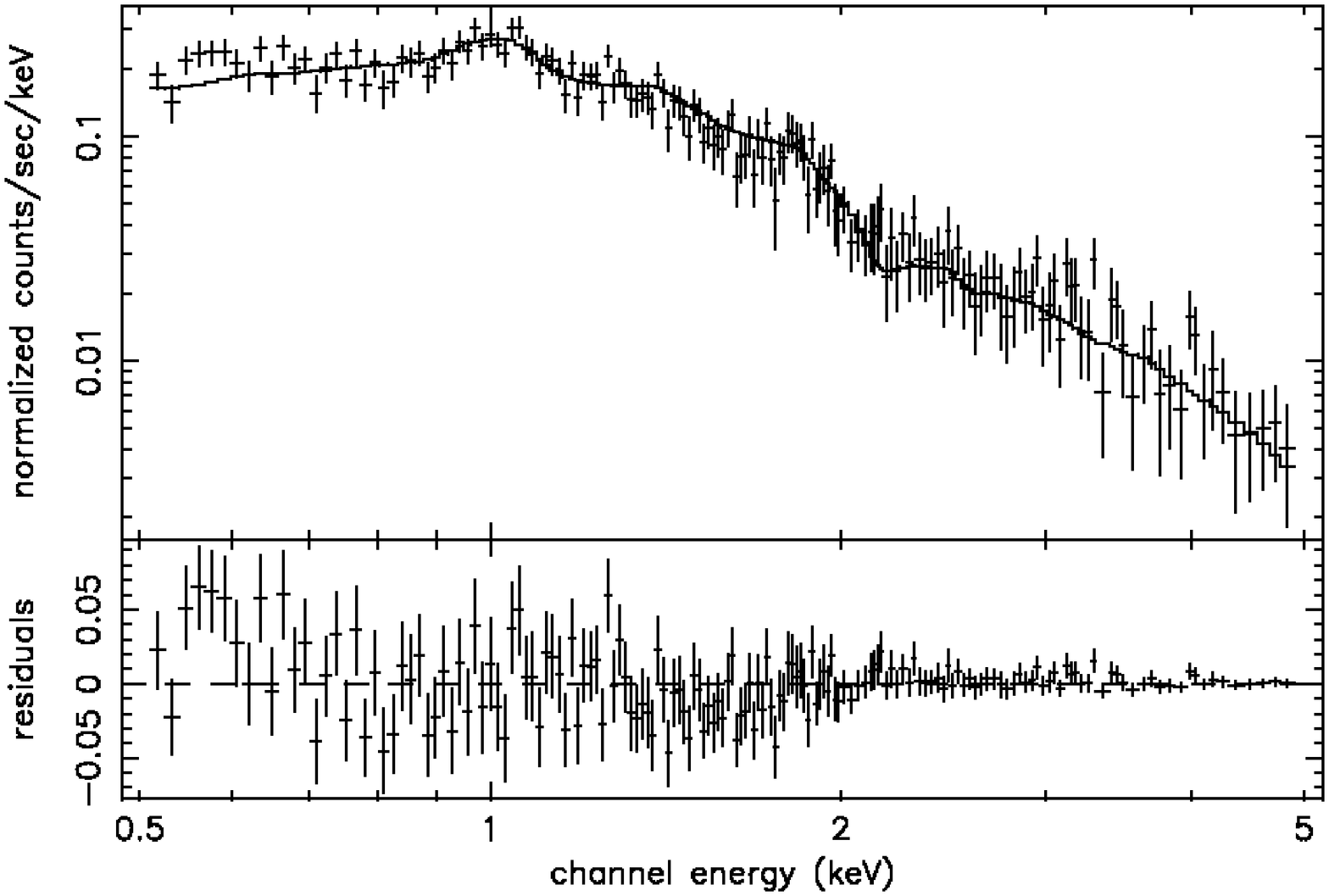}}}
\caption{The spectra of the clusters Abell~160 and Abell~2462.  Both spectra
were fitted with a \MEKAL model.  The emission from the BCG and AGN has been removed.  The lower panel of each plot shows the residuals of the best-fitting model.}
\label{clusspec}
\end{figure*}

The spectra were fitted in the energy range 0.5--5.0\keV with single \MEKAL models modified by the
line-of-sight hydrogen column (\(N_{\mathrm{H}}\)) for each cluster.  The temperatures, metal abundances, (\(Z\)), and \(N_{\mathrm{H}}\) were left free to fit the data resulting in the best-fitting values presented in Table~\ref{clusspecpar}.  No previous X-ray temperature measurement has been made for Abell~160, but the results appear to be in good agreement for other X-ray observations of poor clusters \citep{2003MNRAS.345.1241S}.  The temperature found for Abell~262 is slightly higher than that found from previous work by \citet{Gomez97}, but they fitted a larger cluster area, using significantly fewer counts, which resulted in a poorer fit to their data.  The \(N_{\mathrm{H}}\) values that we measure are consistent with Galactic values (4.3 and 3.1\(\times 10^{20}\)\pcc for Abell~160 and Abell~2462 respectively).  The values of \(Z\) are in agreement with published values of \(Z\) in poor clusters.  The spectra are presented in Fig.~\ref{clusspec} and Table~\ref{clusspecpar}.

\begin{table}
\caption{The best-fitting parameters for the clusters, fitting a \MEKAL
model, and taking into account absorption from the Galaxy.}
\begin{tabular}{llll}
\hline\\ 
Source        & Component      & Parameter              &             \\\hline 
A~160         & \MEKAL         &kT (\keV)               & \(2.7\pm0.2\)\\ 
              &                & \(Z\) (\Zsol) & 0.4\(\pm\)0.1\\
              &                & \(L_X\)                & 4.0\(\pm\)0.02\\
              &                &\(\times\)10\(^{42}\)\ergps&\\
              &                & \(\mathrm{N_H}\) (\(\times 10^{20}\)\pcc)& \(4.4\pm0.1\)\\
              &  \chisq (d.o.f)     &                 & 177 (161) \\\hline
A~2462	      & \MEKAL         & kT (\keV)              & \(2.5\pm0.1\)\\ 
              &                &\(Z\) (\Zsol)   & 0.24\(\pm\)0.05\\
              &                & \(L_X\)    & 2.0\(\pm\)0.1\\
              &                &\(\times\)10\(^{43}\)\ergps &\\ 
              &                &\(\mathrm{N_H}\)\(\times\)10\(^{20}\)\pcc & 3.3\(\pm\)0.2\\
              & \chisq (d.o.f)       &                 & 235 (235)\\\hline
\end{tabular}
\vspace{0.2cm}
\begin{minipage}{7cm}
\small NOTES:\\
Luminosities quoted are absorbed luminosities in the energy range 0.5--5.0\keV\\
\end{minipage}
\label{clusspecpar}
\end{table}

\subsubsection{The WAT host galaxies}
\label{specgals}

We extracted spectra, as described in Section~\ref{speccluster}, from circular regions, centred on the WAT host galaxies, of radius 20 and 10~arcsec for Abell~160 and Abell~2462 respectively.  Background regions were concentric annuli from 20--50~arcsec for Abell~160 and 10--40~arcsec for Abell~2462.

\begin{figure*}
\subfigure[Abell~160]{\scalebox{.3}{\includegraphics[angle=270]{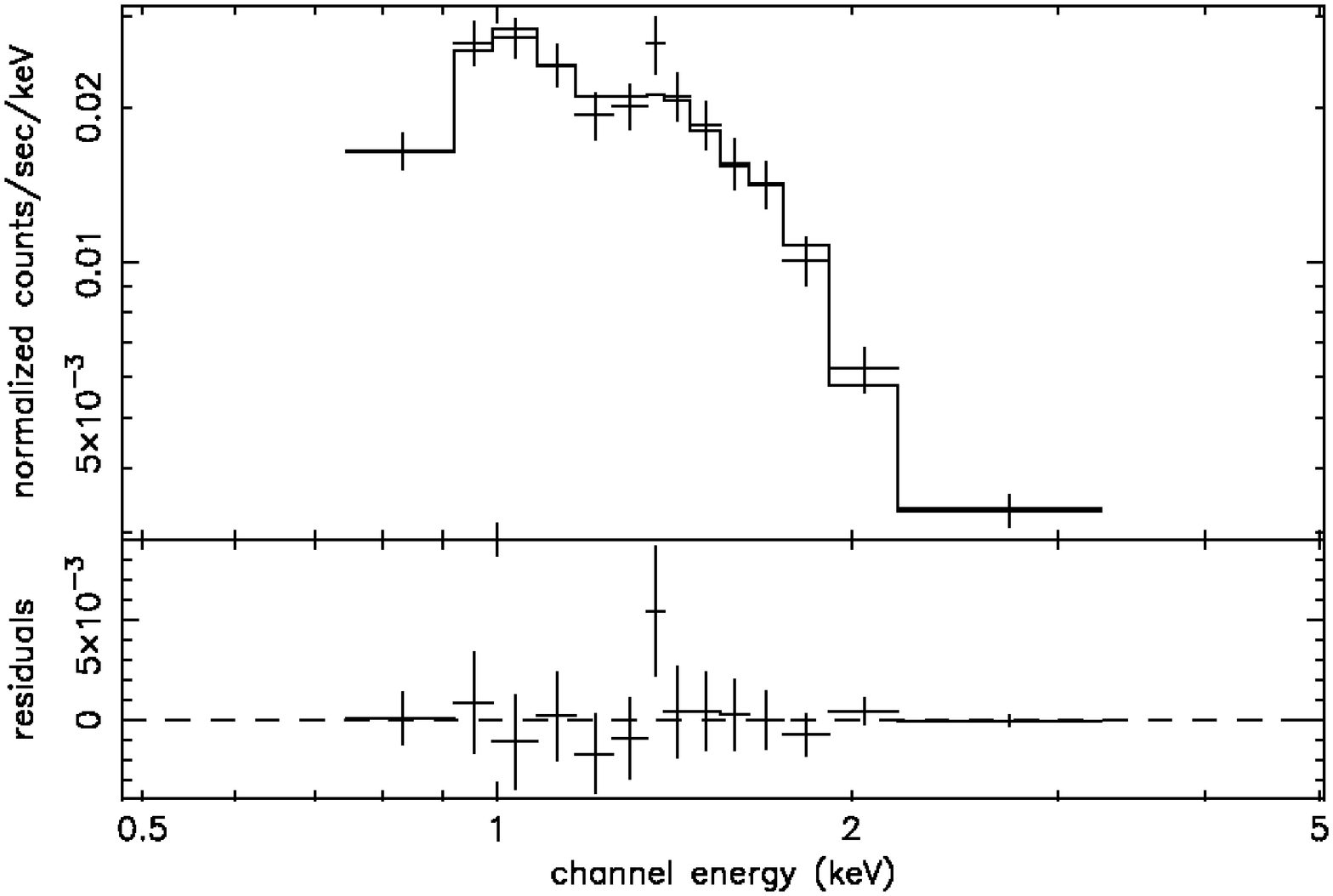}}}
\subfigure[Abell~2462]{\scalebox{.3}{\includegraphics[angle=270]{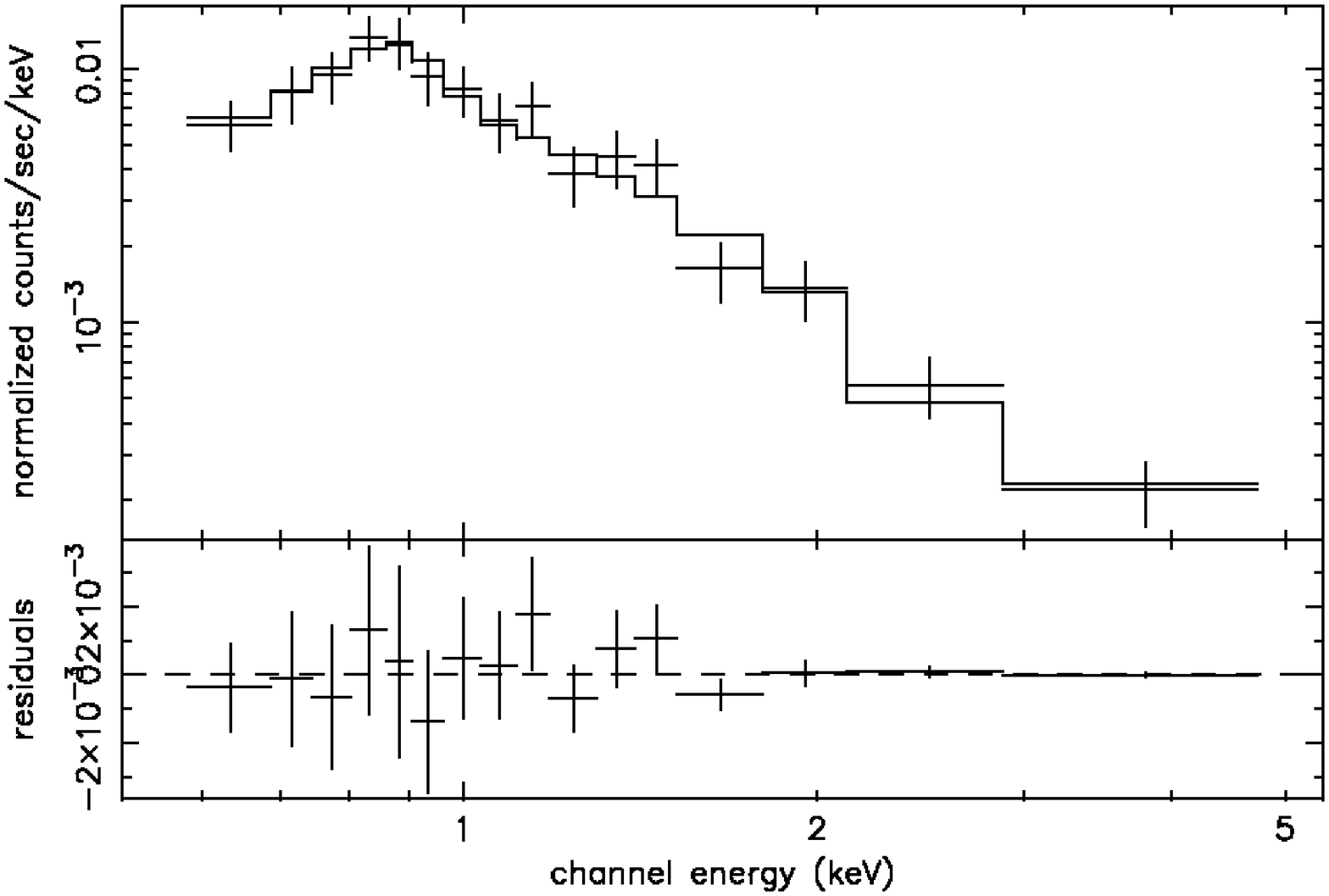}}}
\caption{The power law and \MEKAL fits to the spectra of the BCGs in Abell~160 and Abell~2462.  The solid line shows the best-fitting model, and the residuals to the fit are shown below the spectrum.}
\label{galspec}
\end{figure*}

The galaxy spectra were fitted in {\sc{xspec}} with a composite model consisting of a
\MEKAL model and a power law modified by the line-of-sight hydrogen column. Energy ranges of
0.5\To5.0\keV were used for both BCGs. The spectra and best-fitting
parameters for both BCGs are shown in Fig.~\ref{galspec} and
Table~\ref{spgalpar} respectively.   The index of the power
law model, the temperature of the \MEKAL model, and the normalizations of both models were allowed to vary.  All other
parameters were fixed -- absorption to the Galactic value (given
in Table~\ref{spgalpar}), and metal abundance to 0.3\Zsol.  

Our values for galaxy temperatures appear to be similar to values published by \citet{2002ApJ...569L..79H} for the cD galaxy in Abell~4059 and by \citet{2003ApJ...595..142T} for the cD galaxy in Abell~3112, which are both radio galaxy hosts. 

\begin{table}
\caption{The best-fitting parameters for the galaxies, fitting a \MEKAL
model plus power law.  The absorption was fixed to the Galactic value,
and the metal abundance of the \MEKAL model to 0.3\Zsol.}
\begin{tabular}{llll}
\hline\\ 
Source                   & Component                       & Parameter            & \\
\hline A~160             & \MEKAL                          &kT (\keV)             & \(1.23\pm0.4\)\\
                         &                                 & \(\mathrm{N_H}\) (\pcc)   & 4.3\(\times\)10\(^{20}\)\\ 
                         &Power Law                        & \(\Gamma\)         & 2.22\(\pm0.27\)\\ 
                         &                                 & \(L_X\)            &(1.7\(\pm0.3\mathrm{)}\)\\
                         &                                 & \(\times\)10\(^{41}\)(\ergps)           & \\
                         & \chisq (d.o.f)                 &                    & 16.5 (15) \\\hline
A~2462                   & \MEKAL                          & kT (\keV)          & \(0.79^{+0.2}_{-0.1}\)\\
                         &                                 & \(\mathrm{N_H}\) (\pcc)& 3.11 \(\times\)10\(^{20}\)\\ 
                         &Power Law                        & \(\Gamma\)             & 1.9\(\pm0.2\)\\ 
                         &                                 & \(L_X\)         & (5.8\(\pm0.2\mathrm{)}\)\\
                         &                                 &\(\times\)10\(^{41}\) \ergps & \\
                         & \chisq (d.o.f)                 &        & 10.4 (13) \\\hline

\end{tabular}
\vspace{0.2cm}
\begin{minipage}{7cm}
\small NOTES:\\
Luminosities quoted are absorbed luminosities in the energy range 0.5--5.0\keV\\
\end{minipage}
\label{spgalpar}
\end{table}

\subsubsection{The bright, extended source in Abell~160 -- 'source 1'}
\label{source1}

In Abell~160 there appears to be a bright region of emission near the cluster
redshift (see Section~\ref{xrayobs}).  To investigate what this could be, a spectrum of the
region was extracted using a circular source region of radius 0.5 arcmin (26\kpc)
and a local background in the form of an annulus extending from 0.5 to
1.0 arcmin (53\kpc) both centred on the coordinates stated in Section~\ref{xrayobs}.  We extracted the spectrum in a similar way to that described in Section~\ref{speccluster}.

\begin{figure}
\scalebox{.35}{\includegraphics[angle=270]{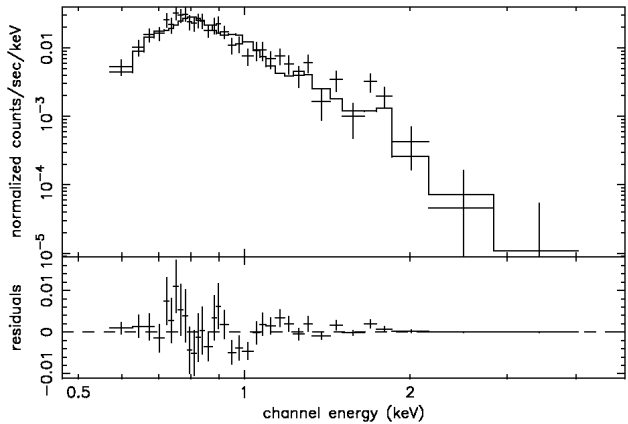}}
\caption{The spectrum and residuals to the best-fitting model of source 1, fitted with a \MEKAL model.}
\label{src1spc}
\end{figure}

This spectrum was fitted with a \MEKAL model, modified by absorption from the line-of-sight hydrogen column.  \(N_{\mathrm{H}}\) was fixed to the Galactic value, and the abundance to 0.3\Zsol.  Allowing the temperature and normalization of the \MEKAL model to vary, we obtained a temperature of 0.52\(^{+0.08}_{-0.06}\)\keV, significantly lower than the temperature of the surrounding ICM.  We present the spectrum of source 1 in Fig.~\ref{src1spc}.

From the DSS optical data, we see that most of the X-ray emission is associated with a large elliptical, at \(z\)=0.0484 \citep{2000AJ....120.2269P} (c.f. \(z\)=0.0447 for the cluster), and there is at least one spectroscopically confirmed `companion' galaxy at a similar redshift \citep[$z$=0.0470;][]{2000AJ....120.2269P}.  Hardness maps of this region (see Section~\ref{deprojanalys} show no evidence of a bow shock in front of the galaxies, and the extended X-ray halo suggests that the galaxies have not been significantly stripped by any encounter with the cluster.  Thus, there appears to be little evidence that this source is interacting with the cluster in any significant way, and we conclude that it is unimportant to the current discussion.

\subsection{Deprojection analysis}
\label{deprojanalys}

As mentioned in Section~\ref{introduction}, it is suggested that the external
environment is a significant factor in causing jets in WATs to flare
suddenly.  Therefore, it is important to determine how the properties of the cluster vary as a function of position.  We constructed temperature and density profiles
of the ICM across the cluster to see if there are any changes in
environment with distance.

\begin{figure*}

\subfigure{\scalebox{.4}{\includegraphics{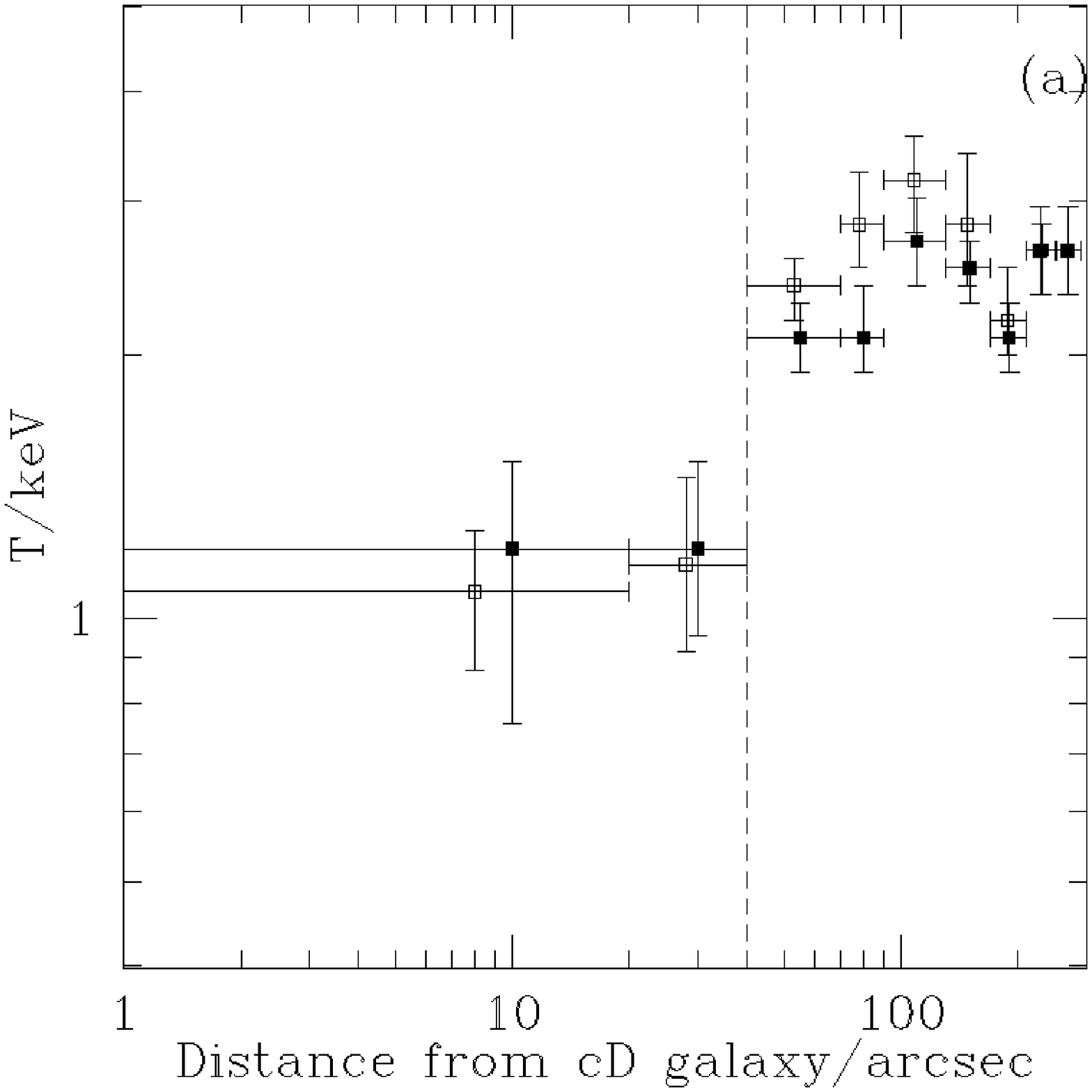}}}
\subfigure{\scalebox{.4}{\includegraphics{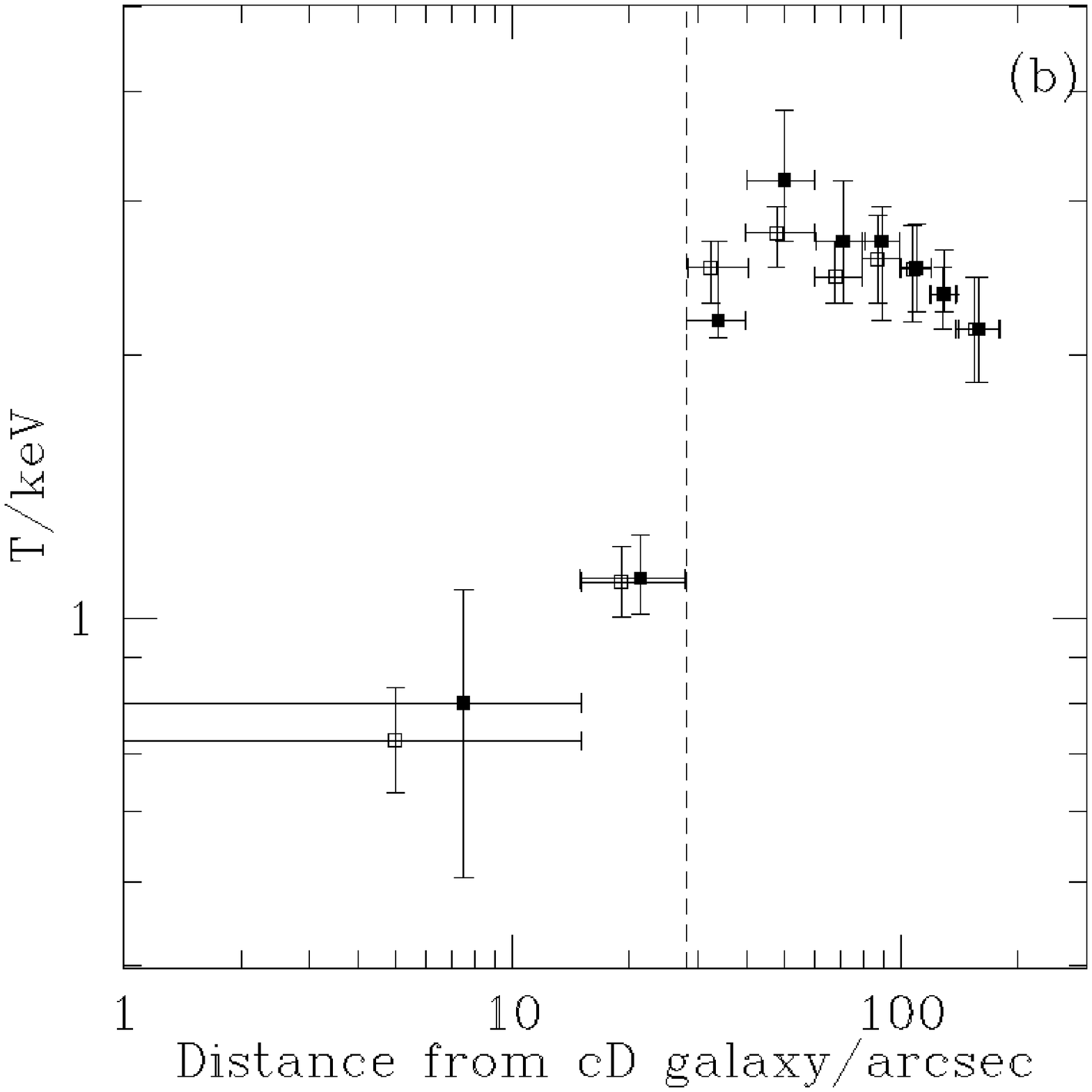}}}
\caption{The temperature profiles centred on the galaxies -- (a) is the profile of Abell~160, and (b) is the profile of Abell~2462.  The metalicities and hydrogen absorbing column densities are set to the values found for the cluster in Section~\ref{specan}.  There appear to be
definite temperature gradients across the galaxy and cluster.  The
closed points are the projected profiles and the open points are the
deprojected profiles.  The dashed vertical lines show the jet
termination point for each radio source.}
\label{tempprof}
\end{figure*}

For each source, spectra were extracted in variable-width annuli
centred on the radio core, such that each annulus contained at least 1000
counts.   In Abell~160, we extracted spectra out to 5.0 arcmin, and in
Abell~2462, out to 3.0 arcmin. For Abell~160, background regions consisted of scaled blank-sky spectra, taken in the same regions as the source spectra.  For Abell~2462, we chose a circular annulus between 3--4~arcmin, and used that as a background for all spectra.  Each spectrum was
then fitted in {\sc{xspec}} with a \MEKAL model, modified by the Galactic absorption.  The metal abundances and Galactic hydrogen column density values were fixed to the values found in Section~\ref{specan}
above.  The projected temperature profiles are shown in
Fig.~\ref{tempprof}.  From the profiles, it is clear that both clusters have cool cores, which are on similar length scales as that of the WAT jets.

\begin{figure}
\center {\scalebox{.5}{\includegraphics{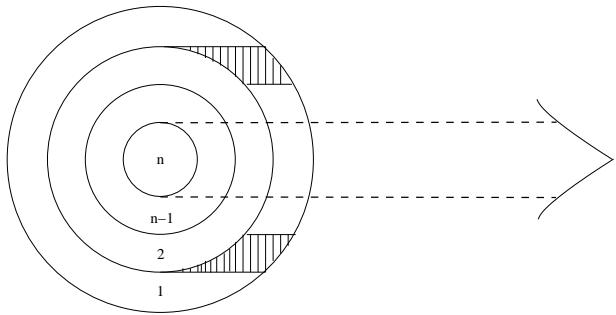}}}
\caption{Schematic diagram of an onion skin deprojection.  We view the
cluster from the right hand side; the dotted lines show the line of
sight.    The cluster is modelled as a set of concentric spheres, labelled 1,2\(\ldots n\) in the diagram.  The hatched area of the first sphere contributes to some of the emission we see from the annulus corresponding to the second sphere.}
\label{deprojfig}
\end{figure}

The data were then deprojected, using an onion skin deprojection
method.  In an onion-skin deprojection, the cluster is modelled as a series of concentric spheres, as showing in Fig.~\ref{deprojfig}.  The outermost sphere is first fitted with a \MEKAL model, with the temperature and normalization free to fit the data.  This determines the temperature and normalization that is appropriate for the outermost sphere (labelled 1 in Fig.~\ref{deprojfig}).  The hatched area of sphere 1 contributes to some of the emission observed from the next annulus, whose radii correspond to those of the sphere 2 in Fig.~\ref{deprojfig}.  The \MEKAL normalization is volume dependent.  The normalization of the first sphere is rescaled to reflect the volume that contributes to the emission from the second sphere.  This first, renormalized \MEKAL component is called component 1.  The second sphere is then fitted with the frozen component 1, and a second \MEKAL component whose temperature and normalization are left free to fit the emission from the second sphere.  This second component is renormalized in a similar way to component 1, and the frozen component 2 and component 1 are then fitted, together with a third \MEKAL component to the third sphere.  This process continues until the (n-1)\(^{th}\) sphere has been fitted.  The innermost sphere is then fitted with the frozen, rescaled, n-1 \MEKAL components, a final \MEKAL component, and in our case, a power law component in order to take account of emission from the active nucleus.  The temperature and normalization of the \MEKAL component, and the normalization of the power law are left free to fit the data.  The power law index was fixed to the value obtained in Section~\ref{specan}.  

The deprojected profile is presented with the projected temperature profiles, in Fig.~\ref{tempprof}.  There appear to be no major differences between the shapes of the two profiles which both agree to within 1\(\sigma\).  Both show a steep temperature gradient in the region where the jet flares, with the profile then showing only a very shallow temperature gradient as we move away from the cluster centre.

The normalization of the \MEKAL model is given by

\begin{equation}
\label{meknorm}
N_{mek}=\frac{10^{-14}}{4\pi \left(D_A\left(1+z\right)\right)^2}\int
n_e n_p dV
\end{equation} where \(N_{mek}\) is the \MEKAL normalization for a particular
annulus, \(D_A\) is the angular diameter distance of the source, and
\(n_e\) and \(n_p\) are the electron and proton densities
respectively.  We further assume that \(n_p=1.18n_e\), and that the
\MEKAL component is fitted in a volume \(V\).  This implies that

\begin{equation}
\label{mekden}
n_e=\left\{\frac{4\pi\left[D_A\left(1+z\right)\right]^2
N_{mek}}{1.18\times10^{-14}V}\right\}^{\frac{1}{2}}
\end{equation}

\begin{figure*}
\subfigure{\scalebox{.35}{\includegraphics{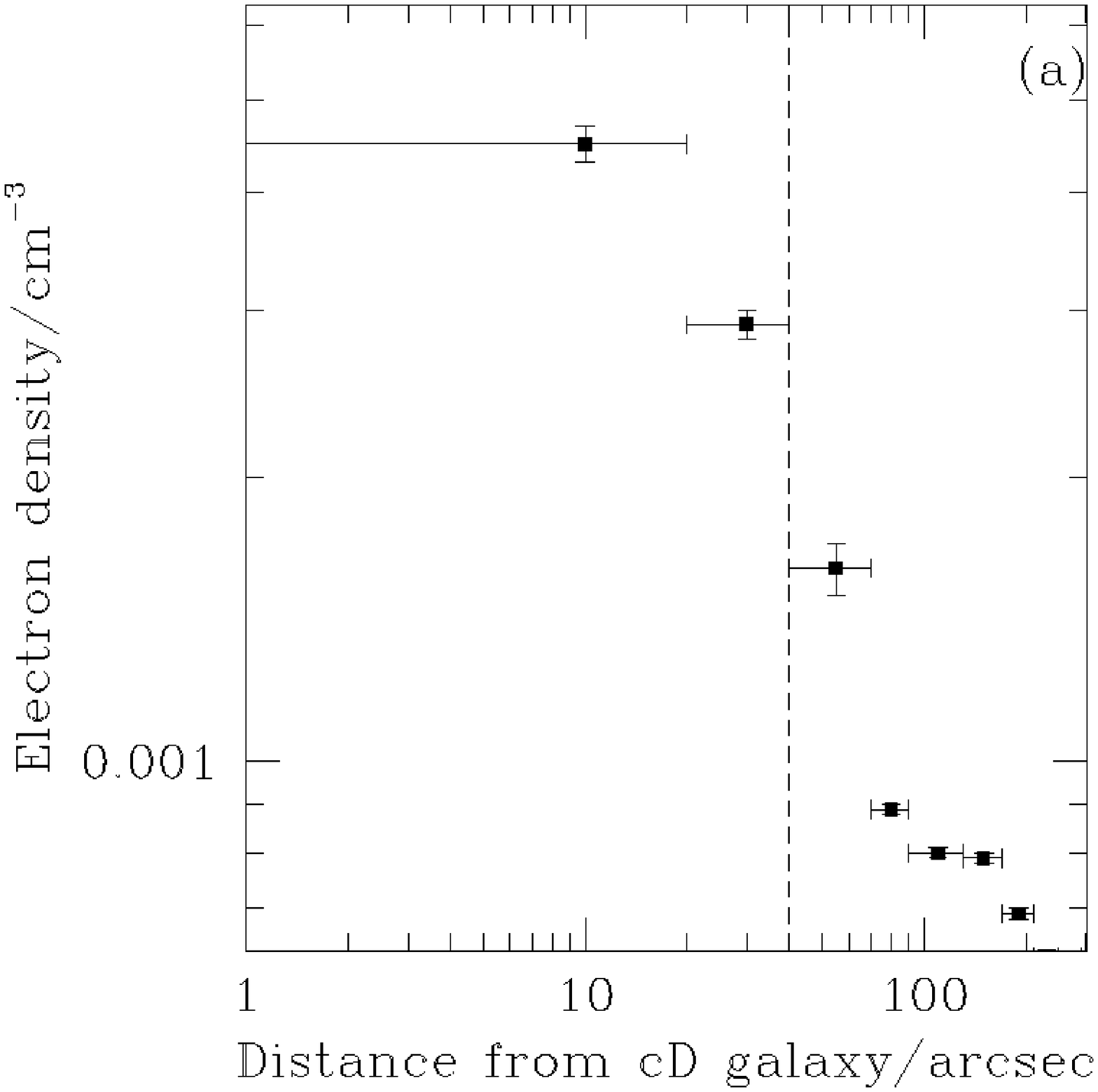}}}
\subfigure{\scalebox{.35}{\includegraphics{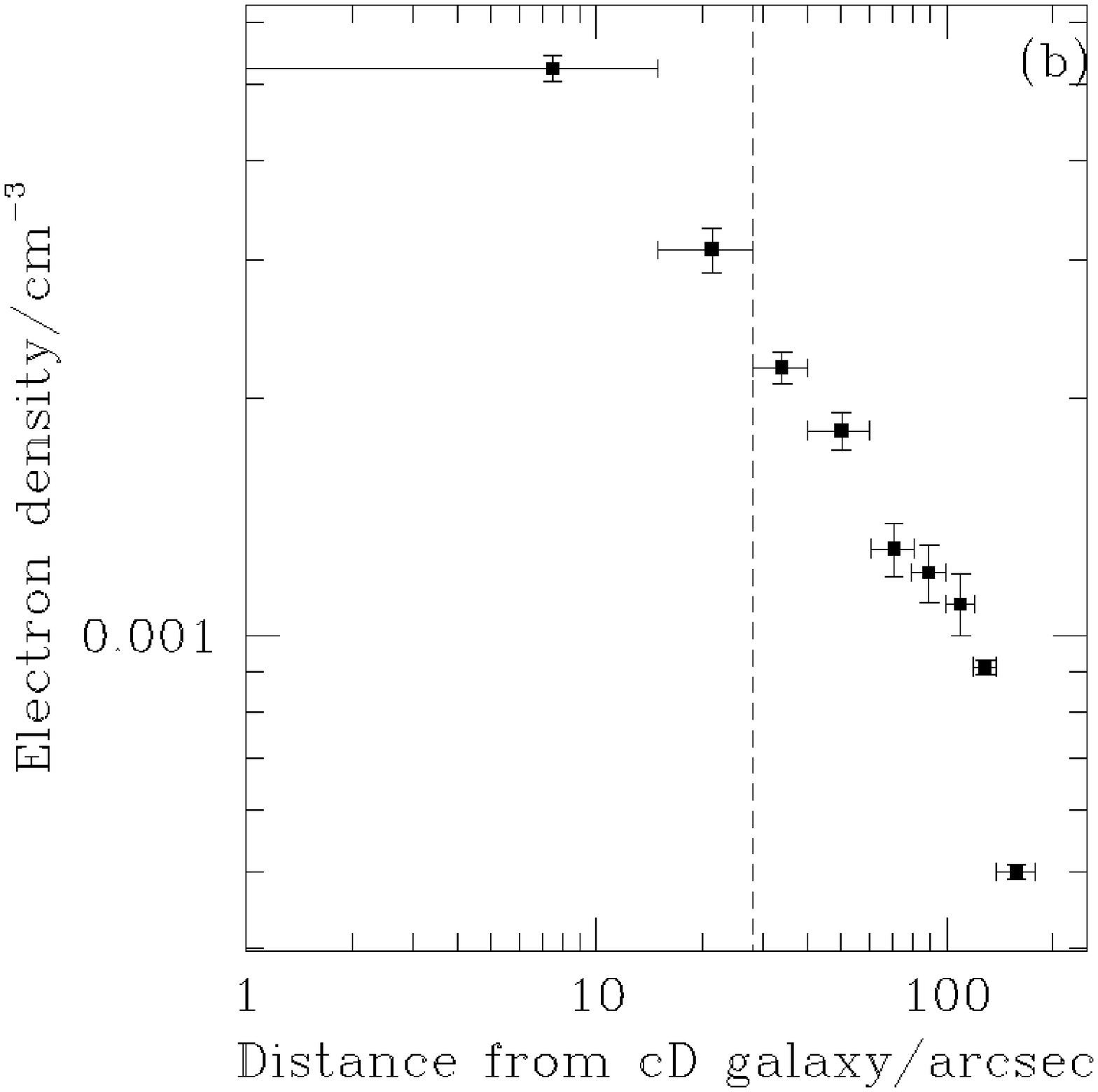}}}
\caption{The deprojected electron density profiles for the two clusters - graph (a) is Abell~160 and graph (b) is Abell~2462.}
\label{densityde}
\end{figure*}

The electron density profiles which are obtained in this manner from the
deprojected temperature profiles are shown in Fig.~\ref{densityde}.
From Fig.~\ref{densityde}, it can be seen in both profiles
that the density decreases with distance from the central galaxy.  In Abell~160, we see a rather steep density gradient over the initial 50~arcsec.  This then flattens out, and decreases at a slower
rate, out to 350~arcsec.  The change in density gradient does not
appear to coincide with the jet flaring point, and in fact occurs
some distance after the point at which the jets flare.  The second cluster, Abell~2462, has a smooth density gradient which
does not show any sudden change; the density drops rather gradually
across the cluster, and the density gradient appears smooth at the jet
flaring point (at 28~arcsec).

This discussion, however, does not take into account projection effects.  We assume here that the radio sources lie on the plane of the sky; however, if they were inclined at some angle \(\theta\) to the line of sight, then we would expect the jets to appear shorter than they actually are.  We can make some estimate of \(\theta\) by considering the properties of the radio jets.  The speed of fluid flow through the jets is thought to be mildly relativistic, with a speed of approximately \(0.3c\) \citep[estimated from jet/counter-jet ratios by][]{HS2003}.  Therefore, we should see the effects of relativistic Doppler boosting in the jets; the jet travelling towards us (the `jet') will be significantly brighter than the jet travelling away from us (the `counter-jet').  Using the radio maps, we find jet/counter-jet ratios of 1.1 for 0110+152 and 2.3 for 2236-176.  These give inclination angles of 3.6 and 33~deg for 0110+152 and 2236-176 respectively, which imply that the deprojected jet lengths would be 40~arcsec for 0110+152 and 33~arcsec for 2236-176.  As expected, the projection effects are not great, since the jets and counter-jets in both sources are of approximately equal brightness, and so projection does not drastically change any of our conclusions here.

\begin{figure*}
\subfigure{\scalebox{.4}{\label{pressure:a}\includegraphics{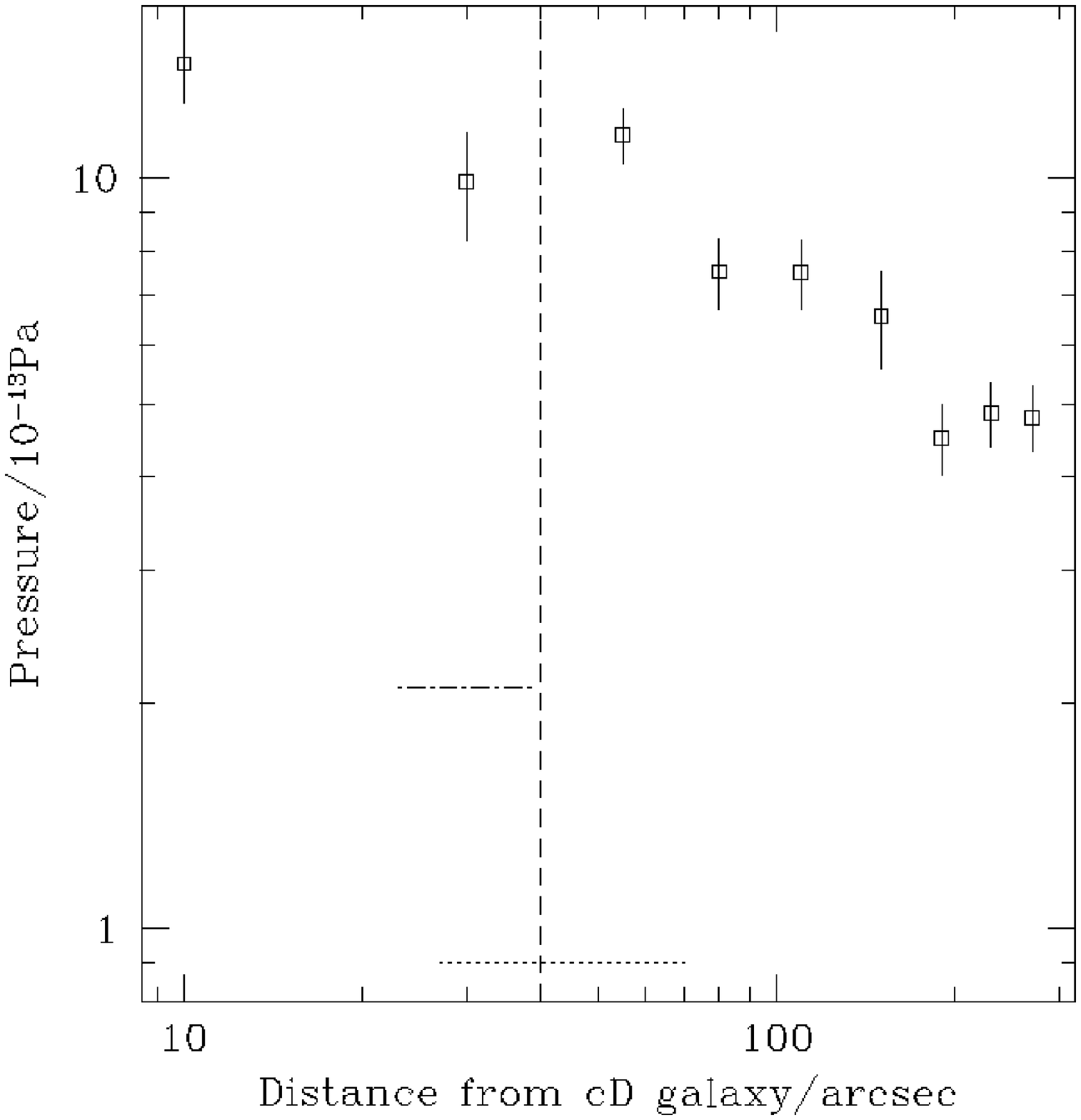}}}
\subfigure{\scalebox{.4}{\label{pressure:b}\includegraphics{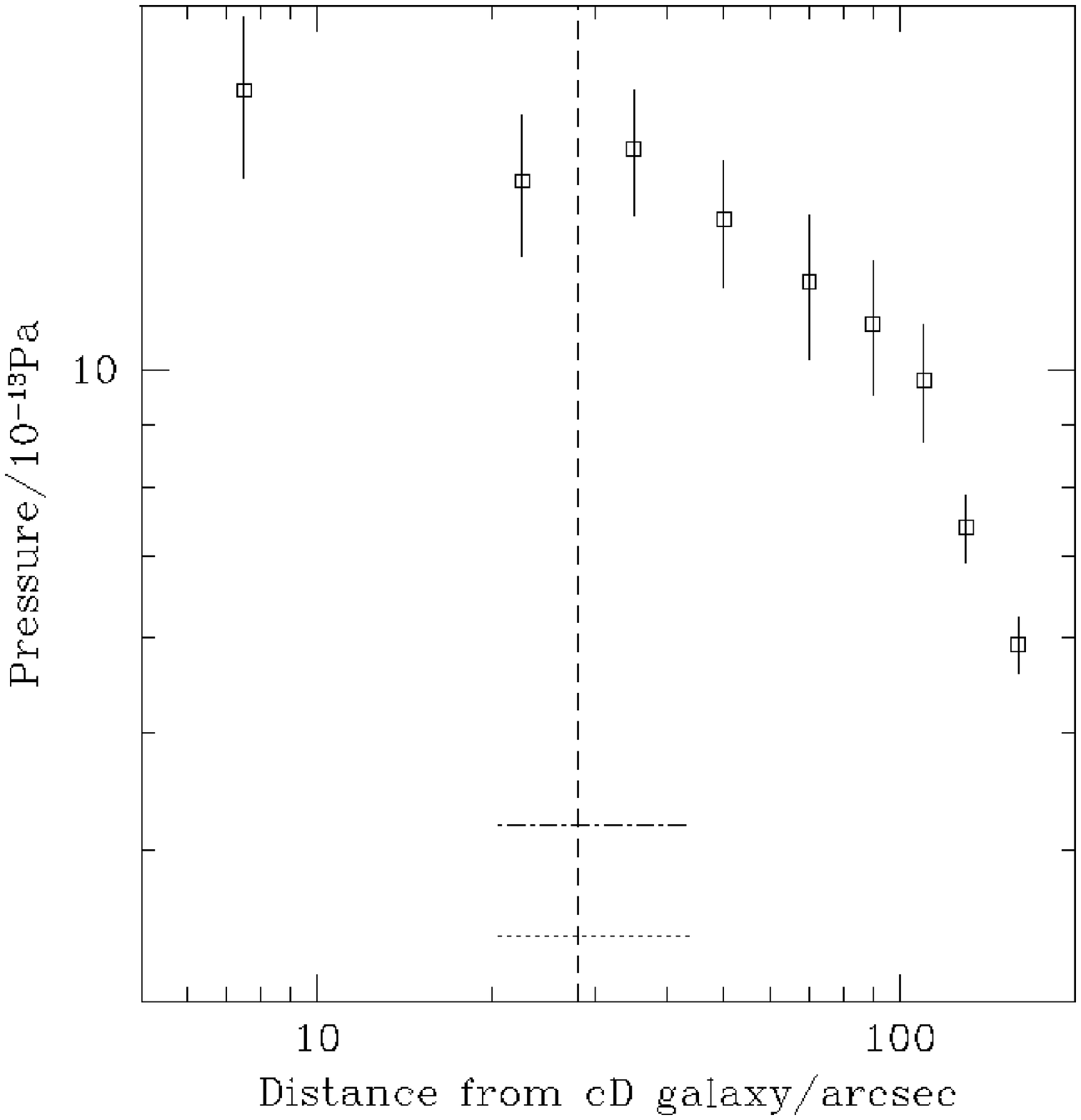}}}
\caption{The pressure profiles for the two clusters (Abell~160 is on the left, and Abell~2462 is on the right).  The vertical
lines show the jet flaring position for each source. The dotted and dash-dotted lines show the minimum pressure, over the distance in which the pressure was calculated, at the base of the North and South lobes of 0110+152 in the plot for Abell~160, and for the East and West lobes in 2236-176 in the Abell~2462 plot.}
\label{pressure}
\end{figure*}

To obtain the pressure profiles, we treat the plasma as an ideal gas
and calculate the pressure using \(P=2.96\times10^{-10}n_ekT\).  Here, \(n_e\) is
the electron number density calculated above in \pcc, \(kT\) is the
temperature of the cluster gas in \keV\(\!\), and we assume that \(n \simeq 1.8n_e\).  The pressure profiles
derived from the deprojected density and temperature profiles are
shown in Fig. \ref{pressure}.  We note that there is a pressure drop
across the cluster, as would be expected.  Also plotted on Fig.~\ref{pressure} are the minimum pressures calculated at the bases of the radio lobes of each source.  We assume power law  electron spectra with electron Lorentz factors between 1 and 2\(\times10^{4}\) with the power-law index of the energy distribution equal to -2.  We also assume that the sources are close to the plane of the sky (as confirmed by our calculations above), and cylindrically symmetric geometry for the plumes.  These minimum pressures are significantly lower than the external thermal pressures, by a factor similar to those seen in the lobes and plumes of more typical FRI sources \citep{2000MNRAS.319..562H}.

The temperature profiles in Fig.~\ref{tempprof}, and the surface
brightness profiles of Fig.~\ref{sbprofiles} are used to define a
galaxy-cluster interface.  From Fig.~\ref{sbprofiles}, it is seen that
galactic emission (at a significantly lower temperature than the
extended emission, as shown by Fig.~\ref{tempprof}) dominates the
brightness profile out to 40 and 20~arcsec for Abell~160 and Abell~2462
respectively.  Fig.~\ref{tempprof} shows that the two innermost annuli
of both clusters have significantly lower temperatures than the outer
bins.  Together, these profiles imply that on a scale of approximately
20\To40~arcsec, there is an `interface' between where the galaxy
dominates, and where the cluster dominates, with a smooth transition
between the two.  

We do not see any evidence for a shock at the jet flaring point; even though the temperature increases rapidly, there is no corresponding jump in the density gradient at that point.  Rather, the density continues to decrease steadily, which is counter-indicative of a shock.  To investigate this in more detail, we also created adaptively binned hardness maps of the central regions of both clusters to search for any variation in hardness that would be indicative of a strong shock that would disrupt the jet.  We used a soft energy range of 0.5--1.5\keV and our `hard' energy band was from 1.5--3.0\keV.  The maps were made using an adaptive binning process detailed in \citet{2001MNRAS.325..178S} to create hardness maps and error maps.  Overlaying contours of the radio sources (Fig.~\ref{radioimages}) onto the temperature maps shows no significant changes in hardness in the regions where the jets flare, compared to regions near to the jet flaring point.  Therefore, it seems that no strong shocks are seen in the data.  Some weak shocks may be present, but it is unlikely that such a weak shock could disrupt jets such as these, which are likely to be moving at speeds much greater than the sound speed in the ICM (\(\sim\)0.3\(c\) for the jet velocity, compared to \(\sim 5\times10^{5}\mathrm{ms^{-1}}\) for the sound speed in the cluster).  

\section{Discussion}
\label{discussion}
From the spatial analysis of the X-ray data (Section \ref{sbanal}),
the jets in both sources flare on similar length scales to that of the
galaxy-cluster transition.  It appears from our analysis of the X-ray data that most existing models fail to account for the jet flaring in WATs.  It is known that the environment must be playing a role in determining the jet properties, but it is not clear how.  Looking at the surface brightness and density profiles (Figs.~\ref{sbprofiles} and ~\ref{densityde}), the density appears to vary smoothly across the point where the jets flare in both sources.  The X-ray gas temperature profiles of both sources show a significant increase in temperature across the point where the jets flare.  We suspect that this is an unresolved steep temperature gradient, particularly as \Chandra observations of the WAT 3C465 \citep[][in prep.]{H3C465} have sufficient signal to noise to detect a temperature gradient in the equivalent region of that source.  In all these sources, and in archival observations of Hydra A (which we have reduced in a similar way to the data presented here) the striking feature is that the transition between the inner, well-collimated jets and the diffuse plumes occurs on a very similar size scale to the transition between a cool, dense `core' at the centre of the cluster and the hotter, more diffuse cluster emission. This leads us to believe that the cool core may be closely related to the location of the jet-plume transition.  We discuss here models put forward to explain the jet flaring process.

\subsection{Models unable to explain the jet flaring}
\label{wrongmodels}

\begin{itemize}

\item The \citet{LRBN95} `cross-wind' model, as detailed in Section~\ref{introduction}, is discussed in \citet{Hardcastle} in relation to 3C130.  The arguments put forward against the cross-wind model in that paper also apply to 0110+152 and 2236-176; bent jets are not seen, and neither do the plumes bend immediately after the jet disruption point, as the cross-wind model would imply.  

\item \citet{LRBN95} also put forward a second model to explain jet disruption -- they model the jet as disrupting when it propagates across a shock in the ICM.  We would expect to see some evidence of shocks in our data, if they are responsible for causing the WAT jets to flare.  However, there is no evidence for a shock in our \Chandra data (see Section~\ref{deprojanalys}), and so this model also fails to explain the jet disruption.  

\item \citet{1999MNRAS.309..273H} model a jet passing through a medium
containing cool dense clouds, which are denser than the ambient medium
by a factor of 50, but are in pressure balance with the ICM.  They
model relativistic jets, in pressure balance with the ICM, hitting
these dense clouds.  The model predicts that moderately fast jets
should disrupt and flare suddenly on coming into contact with a cloud,
and that the radio source should resemble a WAT.  We examined our X-ray images for evidence of cool clouds spatially coincident with the jet flaring points.  In both sources, we find no evidence for any such clouds. Furthermore, finding clouds equidistant from the host galaxy, which disrupt the jets seems rather implausible -- clusters are large when compared to clouds and jets, and having clouds placed equidistant from the host galaxy in all WAT clusters seems too far-fetched a coincidence.

\item \citet{1996ApJ...470..211H} model a jet passing through a power law ISM, and subsequently a constant density ICM.  The jet becomes unstable to the growth of Rayleigh-Taylor and Kelvin-Helmholtz instabilities on the surface of the jet and the jet cocoon upon crossing the ISM/ICM boundary.  Their simulations find that low Mach number jets disrupt on crossing the boundary, and furthermore, disrupt faster than their high Mach number counterparts.  Whilst their simulations produce flaring jets, they assume abrupt gradients in the density and temerapture of the external medium that are not consistent with our observations.  Also, given the best estimates of WAT jet speeds, \citep[approximately \(0.3c\);][]{HS2003}, it is likely that WAT jets have high Mach numbers with respect to the ICM.  To fully understand the processes involved in forming WATs, realistic simulations of radio galaxies in clusters are needed; such simulations need to include models for the ICM that match the data (i.e. continuous density and steep temperature gradients separating the cool core from the rest of the cluster) as well as simulating radio jets with higher Mach numbers.

\end{itemize}

\subsection{An Alternative Model?}
\label{implications}

Up until recently, it has been assumed that the hot X-ray emitting ICM is responsible for the jet termination.  However, as mentioned previously, in almost all cases studied, the jet terminates inside the radio plume.  Therefore, we may have been misled in thinking that the jet termination is determined by the ICM properties.  It might be more appropriate to assume that the the conditions inside the radio plume cause teh jet termination, and the external ICM determines properties such as the shape and length of the radio plumes.

We look to simulations of radio galaxies in clusters for verification of this.  \citet{2004MNRAS.348.1105O} study how radio galaxies in systems with cool cores heat the ICM, and how the radio galaxies themselves develop over time.  Their simulations of a cluster containing a central radio galaxy initially show a source akin to a small FRII, but as time progresses, this radio galaxy gradually starts to look like a WAT.  The jet material then flows back from where the jet terminates towards the cluster centre (their fig.~4).  How far towards the cluster centre this flow extends to must be determined by where the pressure gradient in the cluster halts the jet material.  

\begin{figure}
\scalebox{.4}{\includegraphics{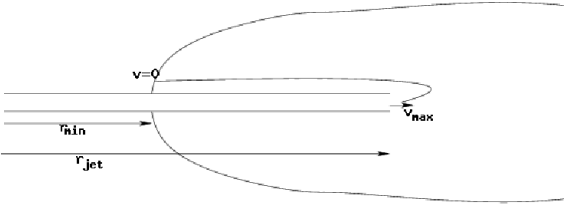}}
\caption{A schematic diagram of what may be happening in a WAT}
\label{backflow}
\end{figure}

We consider the situation in Fig.~\ref{backflow}; material leaves the end of the jet, at a distance \(r_{jet}\), with a velocity \(v_{max}\), and under the influence of gravity flows back up the pressure gradient to the point \(r_{min}\) where it is stopped by the pressure gradient.  This should give us a theoretical estimate as to where the base of the plume should be located.

We consider the radial component of the Navier-Stokes equation:

\begin{equation}
\label{NV}
\rho_{\mathrm{plume}} v \frac{\mathrm{d}v}{\mathrm{d}r} = -\frac{\mathrm{d}P_{\mathrm{plume}}}{\mathrm{d}r} - GM(<r)\frac{\rho_{\mathrm{plume}}}{r^2}
\end{equation} here, \(\rho_{\mathrm{plume}}\) is the density of the radio plume, \(P_{\mathrm{plume}}\) is the internal plume pressure, \(M(<r)\) is the mass of the cluster within a radius \(r\) from the centre of the cluster.  

We then assume that the cluster is in hydrostatic equilibrium, and that the lobe and ICM are in pressure equilibrium, so \(\mathrm{d}P_{\mathrm{plume}}/\mathrm{d}r=\mathrm{d}P_{\mathrm{icm}}/\mathrm{d}r\), and that \(\rho_{\mathrm{plume}}\ll\rho_{\mathrm{icm}}\), and obtain:

\begin{equation}
\label{dif}
\frac{\mathrm{d}v^2}{\mathrm{d}r}\simeq-2\frac{1}{\rho_{\mathrm{plume}}}\frac{\mathrm{d}P_{\mathrm{icm}}}{\mathrm{d}r}
\end{equation} Integrating both sides from where the jet terminates (at \(r_{\mathrm{jet}},\ v=v_{\mathrm{max}}\)) to where \(v=0\) at \(r_{\mathrm{min}}\), we get:

\begin{equation}
\label{result}
v_{\mathrm{max}}^2 \simeq \int^{r_{\mathrm{min}}}_{r_{\mathrm{jet}}} -2 \frac{1}{\rho_{\mathrm{plume}}}\frac{\mathrm{d}P_{\mathrm{icm}}}{\mathrm{d}r} dr
\end{equation}

In order to apply this model to the real data, we can fit the observed densities and pressures, from Figs.~\ref{densityde} and \ref{pressure}, with power laws to simply model the state of the ICM.  Using Abell~160 as an example, we obtain \(P_{\mathrm{icm}}\!\!=\!P_0\left(r/r_{\mathrm{ref}}\right)^{-0.25}\) and \(\rho_{\mathrm{icm}}\!=\!\rho_0\left(r/r_{\mathrm{ref}}\right)^{-0.84}\), where \(P_0\!\!=\!5.35\times10^{-13}\mathrm{Pa}\), and \(\rho_0\!\!=\!4130\,\mathrm{amu}\ \mathrm{m^{-3}}\).  

The distance from the radio core to the base of the plume is \(\sim\)35\kpc\(\ \)in this source.  This can be reproduced from the fits to the data and the above model by taking suitable values for \(v_{\mathrm{max}}\) and \(\rho_{\mathrm{plume}}\), which in this case are \(v_{\mathrm{max}}\!=\!1000\mathrm{kms^{-1}}\) \citep[from][]{2004MNRAS.348.1105O} and \(\rho_{\mathrm{plume}}\!=\!0.001\rho_{\mathrm{icm}}\).  The value for \(\rho_{\mathrm{plume}}\) is, however, several orders of magnitude larger than what we would expect for a fully relativistic plasma, but given that the lobe could be entraining colder gas from its environment, \(\rho_{\mathrm{plume}}\!=\!0.001\rho_{\mathrm{icm}}\) is not entirely unrealistic.

\begin{figure}
\scalebox{.45}{\includegraphics{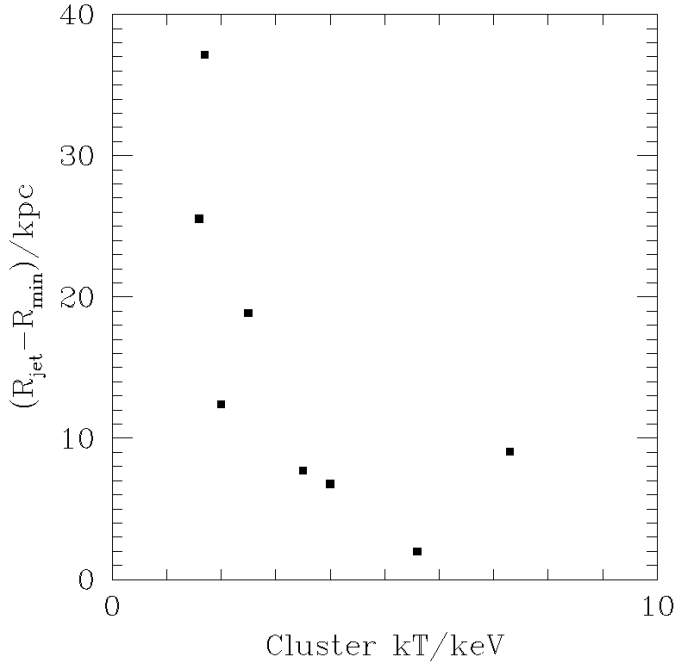}}
\caption{\(r_{\mathrm{jet}}-r_{\mathrm{min}}\) plotted against the host cluster temperature for the \protect{\citet{HS2003}} sample of WATs.  There appears to be an anti-correlation between the cluster temperature and \(r_{\mathrm{jet}}-r_{\mathrm{min}}\)}
\label{rjetplot}
\end{figure}

It appears from the preceeding discussion that we could turn a weak FRII into a WAT by placing it in a cluster with a steep pressure gradient near the core, to surpress the back-flow.  This implies that there should be some sort of relationship between how far upstream the back-flow travels and the temperature of the cluster.  To this end, we use the WAT sample presented in \citet{HS2003}, to determine if there exists any correlation between \(r_{\mathrm{jet}}-r_{\mathrm{min}}\) and the cluster temperature.  We present our findings in Fig.~\ref{rjetplot}, which shows an anti-correlation between  \(r_{\mathrm{jet}}-r_{\mathrm{min}}\) and the cluster temperature.  This suggests that the higher pressures in the centres of hotter clusters may suppress the back-flow.  This may result in conditions in the radio plumes of hotter clusters being altered so that jets in those clusters propagate through smaller distances in the radio plume before being disrupted.  

From our results, it appears that a cool core, joined to the
rest of the cluster by a region with a steep temperature gradient, is
an essential feature of any WAT host cluster.  It is not clear
how such a cool component might cause a WAT to form from a weak FRII. One
possibility is that the link arises from the higher pressure within cool
cores. From hydrostatic equilibrium it can be seen that for a given value of
\(M(<r)/r^2\), the pressure gradient is steeper when
\(\rho_{\mathrm{icm}}\) is larger. Hence, in dense cool cores, the steeper
pressure gradient could inhibit back-flow more effectively,
giving a plume whose base is closer to the jet termination
point.

\citet{HS2003} also find a significant anti-correlation
between \(r_{\mathrm{jet}}\) and the cluster temperature -- jets in hotter
clusters are shorter than those in cooler clusters.  This suggests that
the cluster is a factor in determining the location of the jet
termination. The trend for shorter jets in hotter systems may be related to their higher
gas pressures, but simulations of weak FRIIs in clusters that contain
realistic cool cores are needed to investigate this.

\section{Concluding Remarks}
\label{conclusions}

We find that new \Chandra data shows a relationship between the radio morphology of WATs, specifically the jet length, and the temperature distribution of the cluster medium.  In particular, it appears that the location of the base of the plume is correlated with the steep temperature gradient that separates a cool core from the rest of the ICM.  However, we are still unclear on the mechanism behind jet flaring and the locating of the base of the plume.  Firstly, we need to establish whether a cool core on scales similar to the length of the radio jets is the exception rather than the rule; secondly we need to understand why the steep temperature or pressure gradient causes jets to flare.  The first point requires deeper X-ray observations of a large, statistically significant sample of WAT host clusters to compare with both clusters that host radio quiet galaxies, and clusters that host `normal' FRI galaxies.  For the second point, the properties of WAT jets and of their cluster environments need to be fed into realistic cluster models in order to establish whether the steep temperature gradient is in itself responsible for the jet flaring, or rather is simply a tracer of some underlying phenomenon that causes the jets to make the sudden transition to a diffuse plume.

\section*{acknowledgements}

NNJ thanks PPARC for a research studentship.  MJH thanks the Royal Society for a research fellowship.  The National Radio Astronomy Observatory is a facility of the National Science Foundation operated under cooperative agreement by Associated Universities, Inc.

\bibliographystyle{mn2e}
\bibliography{ME1427rv_bibliography}

\begin{thebibliography}{}

\bibitem[\protect\citeauthoryear{{Acreman} et~al.,}{{Acreman}
  et~al.}{2005}]{Acreman}
{Acreman} D.,  et~al., 2005, In prep.

\bibitem[\protect\citeauthoryear{{Begelman}, {Rees} \& {Blandford}}{{Begelman}
  et~al.}{1979}]{BRB79}
{Begelman} M.~C.,  {Rees} M.~J.,    {Blandford} R.~D.,  1979, Nature, 279, 770

\bibitem[\protect\citeauthoryear{{Cavaliere} \& {Fusco-Femiano}}{{Cavaliere} \&
  {Fusco-Femiano}}{1976}]{KFF76}
{Cavaliere} A.,  {Fusco-Femiano} R.,  1976, \aap, 49, 137

\bibitem[\protect\citeauthoryear{{Drake}, {Merrifield}, {Sakelliou} \&
  {Pinkney}}{{Drake} et~al.}{2000}]{Drake}
{Drake} N.,  {Merrifield} M.~R.,  {Sakelliou} I.,    {Pinkney} J.~C.,  2000,
  \mnras, 314, 768

\bibitem[\protect\citeauthoryear{{Fanaroff} \& {Riley}}{{Fanaroff} \&
  {Riley}}{1974}]{fr1974}
{Fanaroff} B.~L.,  {Riley} J.~M.,  1974, {\mnras}, 167, 31

\bibitem[\protect\citeauthoryear{{G\'{o}mez}, {Pinkney}, {Burns}, {Wang},
  {Owen} \& {Voges}}{{G\'{o}mez} et~al.}{1997}]{Gomez97}
{G\'{o}mez} P.~L.,  {Pinkney} J.,  {Burns} J.~O.,  {Wang} Q.,  {Owen} F.~N.,
  {Voges} W.,  1997, \apj, 474

\bibitem[\protect\citeauthoryear{Hardcastle}{Hardcastle}{1998}]{Hardcastle}
Hardcastle M.,  1998, \mnras, 298, 569

\bibitem[\protect\citeauthoryear{Hardcastle \& Sakelliou}{Hardcastle \&
  Sakelliou}{2004}]{HS2003}
Hardcastle M.,  Sakelliou I.,  2004, \mnras

\bibitem[\protect\citeauthoryear{{Hardcastle} \& {Worrall}}{{Hardcastle} \&
  {Worrall}}{2000}]{2000MNRAS.319..562H}
{Hardcastle} M.~J.,  {Worrall} D.~M.,  2000, \mnras, 319, 562

\bibitem[\protect\citeauthoryear{{Hardcastle et al}}{{Hardcastle et
  al}}{2005}]{H3C465}
{Hardcastle et al} 2005, In prep.

\bibitem[\protect\citeauthoryear{{Heinz}, {Choi}, {Reynolds} \&
  {Begelman}}{{Heinz} et~al.}{2002}]{2002ApJ...569L..79H}
{Heinz} S.,  {Choi} Y.,  {Reynolds} C.~S.,    {Begelman} M.~C.,  2002, \apjl,
  569, L79

\bibitem[\protect\citeauthoryear{{Higgins}, {O'Brien} \& {Dunlop}}{{Higgins}
  et~al.}{1999}]{1999MNRAS.309..273H}
{Higgins} S.~W.,  {O'Brien} T.~J.,    {Dunlop} J.~S.,  1999, \mnras, 309, 273

\bibitem[\protect\citeauthoryear{{Hooda} \& {Wiita}}{{Hooda} \&
  {Wiita}}{1996}]{1996ApJ...470..211H}
{Hooda} J.~S.,  {Wiita} P.~J.,  1996, \apj, 470, 211

\bibitem[\protect\citeauthoryear{{Jones} \& {Forman}}{{Jones} \&
  {Forman}}{1999}]{1999ApJ...511...65J}
{Jones} C.,  {Forman} W.,  1999, \apj, 511, 65

\bibitem[\protect\citeauthoryear{Leahy}{Leahy}{1993}]{Leahy93}
Leahy J.,  1993, Jets in Extragalactic Radio Sources.
Springer-Verlag, Heidelberg

\bibitem[\protect\citeauthoryear{{Loken}, {Roettiger}, {Burns} \&
  {Norman}}{{Loken} et~al.}{1995}]{LRBN95}
{Loken} C.,  {Roettiger} K.,  {Burns} J.~O.,    {Norman} M.,  1995, \apj, 445,
  80

\bibitem[\protect\citeauthoryear{{O'Donoghue}, {Eilek} \& {Owen}}{{O'Donoghue}
  et~al.}{1990}]{ODonoghue}
{O'Donoghue} A.~A.,  {Eilek} J.~A.,    {Owen} F.~N.,  1990, \apjs, 72, 75

\bibitem[\protect\citeauthoryear{{O'Donoghue}, {Eilek} \& {Owen}}{{O'Donoghue}
  et~al.}{1993}]{1993ApJ...408..428O}
{O'Donoghue} A.~A.,  {Eilek} J.~A.,    {Owen} F.~N.,  1993, \apj, 408, 428

\bibitem[\protect\citeauthoryear{{Omma}, {Binney}, {Bryan} \& {Slyz}}{{Omma}
  et~al.}{2004}]{2004MNRAS.348.1105O}
{Omma} H.,  {Binney} J.,  {Bryan} G.,    {Slyz} A.,  2004, \mnras, 348, 1105

\bibitem[\protect\citeauthoryear{{Pinkney}, {Burns} \& {Hill}}{{Pinkney}
  et~al.}{1994}]{1994AJ....108.2031P}
{Pinkney} J.,  {Burns} J.~O.,    {Hill} J.~M.,  1994, \aj, 108, 2031

\bibitem[\protect\citeauthoryear{{Pinkney}, {Burns}, {Ledlow}, {G{\' o}mez} \&
  {Hill}}{{Pinkney} et~al.}{2000}]{2000AJ....120.2269P}
{Pinkney} J.,  {Burns} J.~O.,  {Ledlow} M.~J.,  {G{\' o}mez} P.~L.,    {Hill}
  J.~M.,  2000, \aj, 120, 2269

\bibitem[\protect\citeauthoryear{{Sanders} \& {Fabian}}{{Sanders} \&
  {Fabian}}{2001}]{2001MNRAS.325..178S}
{Sanders} J.~S.,  {Fabian} A.~C.,  2001, \mnras, 325, 178

\bibitem[\protect\citeauthoryear{{Sanderson} \& {Ponman}}{{Sanderson} \&
  {Ponman}}{2003}]{2003MNRAS.345.1241S}
{Sanderson} A.~J.~R.,  {Ponman} T.~J.,  2003, \mnras, 345, 1241

\bibitem[\protect\citeauthoryear{{Sarazin}}{{Sarazin}}{1988}]{Sarazin}
{Sarazin} C.~L.,  1988, {X-ray emission from clusters of galaxies}.
Cambridge Astrophysics Series, Cambridge: Cambridge University Press, 1988

\bibitem[\protect\citeauthoryear{{Takizawa}, {Sarazin}, {Blanton} \&
  {Taylor}}{{Takizawa} et~al.}{2003}]{2003ApJ...595..142T}
{Takizawa} M.,  {Sarazin} C.~L.,  {Blanton} E.~L.,    {Taylor} G.~B.,  2003,
  \apj, 595, 142

\bibitem[\protect\citeauthoryear{{Worrall}, {Birkinshaw} \&
  {Hardcastle}}{{Worrall} et~al.}{2001}]{WBH2001}
{Worrall} D.~M.,  {Birkinshaw} M.,    {Hardcastle} M.~J.,  2001, \mnras, 326,
  L7

\end{thebibliography}

\end{document}